\documentclass{article}

\usepackage[accepted]{icml2025}

\usepackage{natbib}
\usepackage{shortex}

\icmltitlerunning{AutoStep: Locally adaptive involutive MCMC}

\begin{document}

\twocolumn[
\icmltitle{AutoStep: Locally adaptive involutive MCMC}

\icmlsetsymbol{equal}{*}

\begin{icmlauthorlist}
\icmlauthor{Tiange Liu}{ubc}
\icmlauthor{Nikola Surjanovic}{ubc}
\icmlauthor{Miguel Biron-Lattes}{ubc}
\icmlauthor{Alexandre Bouchard-C{\^o}t{\'e}}{ubc}
\icmlauthor{Trevor Campbell}{ubc}
\end{icmlauthorlist}

\icmlaffiliation{ubc}{Department of Statistics, University of British Columbia, Canada}
\icmlcorrespondingauthor{Trevor Campbell}{trevor@stat.ubc.ca}
\icmlkeywords{Markov chain Monte Carlo, locally adaptive, involution, step size, tuning}

\vskip 0.3in
]

\printAffiliationsAndNotice{}

\begin{abstract}
Many common Markov chain Monte Carlo (MCMC) kernels can be formulated using a
deterministic involutive proposal with a step size parameter.  Selecting an
appropriate step size is often a challenging task in practice; and for complex
multiscale targets, there may not be one choice of step size that works well
globally.  In this work, we address this problem with a novel class of
involutive MCMC methods---AutoStep MCMC---that selects an appropriate step size
at each iteration adapted to the local geometry of the target distribution.  We
prove that under mild conditions AutoStep MCMC is $\pi$-invariant, irreducible, and aperiodic, 
and obtain bounds on expected energy jump distance and cost per iteration.
Empirical results examine the robustness and efficacy of our proposed step size
selection procedure, and show that AutoStep MCMC is competitive with
state-of-the-art methods in terms of effective sample size per unit cost on a
range of challenging target distributions.
\end{abstract}

\section{Introduction}
\label{sec:intro} 

Markov chain Monte Carlo (MCMC) \citep{Metropolis53, Hastings70} is an effective tool for
approximating integrals arising in Bayesian inference
problems. The performance of MCMC is often sensitive to the choice of tuning
parameters in the Markov kernel. In particular, methods
that propose a new state followed by an accept/reject step---e.g., random-walk
Metropolis--Hastings (RWMH) \citep{Hastings70}, the Metropolis-adjusted Langevin 
algorithm (MALA) \citep{Rossky78}, and Hamiltonian Monte Carlo (HMC) \citep{Duane87,Neal96}---often 
involve a scalar \emph{step size} parameter $\theta \geq 0$ that 
governs the distance of the proposed next state from the current state.  Too large a 
choice of $\theta$ results in distant proposals that are often rejected, while too
small a choice leads to nearby proposals that do not explore the state space
quickly; either case results in slow convergence of the chain. For certain 
multiscale targets (e.g. Bayesian posteriors with scale priors \citep{PolsonScott12}) 
there may not even be a single good choice of step size
throughout the whole state space.

Existing methods for selecting step size parameters fall generally into three
categories: adaptive MCMC, discrepancy minimization, and locally-adaptive kernels.
Adaptive MCMC algorithms \citep{Haario01,Atchade06,Andrieu08,Marshall12} 
tune the proposal distribution using previous draws from the chain,
often targeting a particular acceptance rate derived from high-dimensional asymptotics \citep{Roberts97,Roberts98}.
Obtaining theoretical guarantees on estimates produced 
by adaptive MCMC targeting a distribution $\pi$ 
is technically difficult and often
requires strict conditions on the adaptation process, such as increasingly infrequent adaptation \citep{Chimisov18}.
Discrepancy minimization \citep{neklyudov_metropolis-hastings_2018,Coullon23} 
involves tuning using a divergence between the empirical distribution of draws and the target $\pi$, 
which requires multiple MCMC runs to estimate the divergence for each candidate step size.
Both approaches also identify only a single step size value, which may not be appropriate for the whole state space.

Locally-adaptive kernels, in contrast, select a value for the step size
\emph{at each iteration} based on the current state. 
Because the step size depends on the current state, these kernels can adapt to the local
shape of the target $\pi$; and because they depend only on the current state,
they are Markovian and one can use standard tools to prove $\pi$-invariance.
There are many approaches to locally adaptive step size selection in the literature.
Mixture kernels with state-dependent weights \citep{Maire22} and delayed-rejection \citep{Tierney99,Mira01}
are both general approaches, but each requires a prespecified maximum number of step sizes to consider at each iteration.
There are also numerous methods specific 
to certain samplers, e.g., HMC and MALA \citep{Girolami11,Nishimura16,Kleppe16,Modi24,BironLattes24,Turok24},
RWMH \citep[see][]{Livingstone21}, or slice sampling \citep{Neal03}.
Of these, the method most related to the present work is autoMALA \citep{BironLattes24},
which chooses a step size in MALA using a doubling/halving procedure that targets 
a Metropolis--Hastings acceptance ratio in a randomized range 
$(a, b)\subset [0,1]$. While the method
was shown to be $\pi$-invariant, it was crucially not shown to be either irreducible
or aperiodic. In practice, AutoMALA indeed can get stuck in certain states,
especially when the initialization point is far from the mass of the target distribution, as is commonly the 
case in Bayesian inference; to address this, AutoMALA 
requires inexact unadjusted steps (see Algorithm 3 in \citet{BironLattes24}).

In this work, we develop a novel method, \emph{AutoStep MCMC}, for locally-adaptive step size selection in
the broad class of involutive MCMC methods \citep{tierney1998,andrieu_general_2020,Neklyudov20}.
We show that these Markov kernels are $\pi$-invariant, irreducible, and aperiodic
under mild assumptions on the target distribution, thereby substantially generalizing
and improving upon AutoMALA. We further provide bounds
on energy jump distances and expected cost per iteration, demonstrating the robustness
of the method to the setting of an initial step size parameter. 
Empirical results confirm the theory and demonstrate that AutoStep
is stable, reliable, and competitive with other adaptive 
methods.
Proofs of all theoretical results are provided in \cref{sec:proofs}.

Note that concurrent work on Gibbs self-tuning \citep{BouRabee24,BouRabee24NUTS}
introduces the same general technique for locally-adaptive involutive MCMC, proves $\pi$-invariance, and develops 
specific adaptation schemes for HMC and NUTS \citep{Neal11,HoffmanGelman14}. In this work,
we provide an analysis of $\pi$-invariance, irreducibility, aperiodicity, 
and expected cost and performance, with a specific focus 
on involutions parametrized by a step size. 

\section{Background}
\label{sec:background}

Let $\pi$ be a given target probability distribution on
an open subset $\scX \subset \reals^d$. With a slight abuse of notation, we assume that $\pi$
admits a density $\pi(x)$ with respect to the Lebesgue measure on $\reals^d$, 
and that we can evaluate a function $\gamma(x) \propto \pi(x)$ pointwise so
that 
\[
   \pi(x) = \frac{\gamma(x)}{\int \gamma(x)\d x},
\]
where $\int \gamma(x)\d x$ is the unknown normalizing constant.

Involutive Markov Chain Monte Carlo \citep{tierney1998,andrieu_general_2020,Neklyudov20} 
is an MCMC method that uses \emph{involutions}, i.e., functions $f$ where $f^{-1}=f$, 
to generate new proposals.  While there are many
possible variations of involutive MCMC, in this work we use the following
formulation. Fix a distribution $m$ on an open subset $\scZ \subset \reals^p$ with density
$m(z)$ with respect to the Lebesgue measure, and a family of
differentiable involutions $f_\theta: \scX\times\scZ \to
\scX\times \scZ$ parametrized by $\theta\in\Theta$. Then, starting from a state
$x_t$, we draw $z_t \dist m$ and the proposal
\[
   \label{eq:basekernel1}
   x'_{t+1}, z'_{t+1} &= f_\theta(x_t, z_t).
\]
We set the next state to $x_{t+1} = x'_{t+1}$ with probability
\[
   \min\lt\{1, \exp(\ell(x_t,z_t,\theta))\rt\}, \label{eq:basekernel2}
\]
where
\[
   \ell(x_t,z_t,\theta) = \log\lt(\frac{\pi(x'_{t+1})m(z'_{t+1})}{\pi(x_t)m(z_t)}\lt|\nabla f_\theta(x_t,z_t) \rt|\rt),
   \label{eq:elldefn}
\]
and otherwise set it to $x_{t+1} = x_t$. The sequence $x_t$ is a Markov chain
and has stationary distribution $\pi$ if $f_\theta$ is continuously
differentiable \citep[Thm.~2]{tierney1998}.  

Choosing different families of involutions $\{f_\theta\}$ and auxiliary distributions $m$ yields different
MCMC algorithms. For example, random walk Metropolis--Hastings
(RWMH) \citep{Hastings70} with step size $\theta > 0$ and 
mass matrix $M$ is obtained by setting
\[
f_\theta(x,z) &= (x+\theta M^{-1}z, -z) &  m &= \Norm(0, M).\label{eq:rwmh}
\]
The Metropolis-adjusted Langevin algorithm (MALA) \citep{Rossky78} with step size $\theta > 0$
and mass matrix $M$ is obtained by setting
\[
f_\theta(x,z) &= (x', -z')  & m &= \Norm(0,M), \label{eq:mala}
\]
where $x', z'$ are computed via the \emph{leapfrog} map
\[
\begin{aligned}
   z_{1/2} & \gets z + \frac{\theta}{2} \nabla \log \pi(x) \\
   x' & \gets x + \theta M^{-1}z_{1/2}\\
   z' & \gets z_{1/2} + \frac{\theta}{2} \nabla \log \pi(x').
\end{aligned}\label{eq:leapfrog}
\]
Finally, Hamiltonian Monte Carlo (HMC) \citep{Duane87,Neal96} with step size 
$\theta > 0$, mass matrix $M$, and path length $L$ is obtained by setting
\[
f_\theta(x,z) &= (x', -z') & m &= \Norm(0, M), \label{eq:hmc}
\]
where $x', z'$ are computed via $L$ leapfrogs.

Many involutive MCMC methods---including the above three examples---have 
a positive scalar tuning parameter $\theta > 0$ that can be interpreted 
as a form of ``step size'': larger values result in more distant
proposals, while smaller values result in nearby proposals. Too large
a choice of $\theta$ results in many rejected proposals, while too small a choice
results in proposals that are accepted but explore the state space slowly.
Furthermore, there may not be a single choice of $\theta$ that applies globally,
e.g., in the case of multiscale targets \citep{PolsonScott12}. 
This work resolves this problem by selecting an appropriate $\theta$ at each
iteration depending on the local behaviour of the augmented target $\pi\cdot m$.

\section{AutoStep MCMC}
\label{sec:automcmc}

In this section, we develop \emph{AutoStep MCMC}, a family of
modified involutive MCMC methods that automatically select appropriate
tuning parameter values at each iteration.
The key technique in developing AutoStep MCMC is to formulate the sampler on
an augmented space that includes the tuning parameter $\theta \in\Theta$ as well as other
auxiliary quantities.
For a given family of continuously 
differentiable involutions $\{f_\theta : \theta\in\Theta\}$ 
on $\scX\times \scZ$, define the augmented space $\scS$ as
\[
\scS = \scX \times \scZ \times \Delta \times \Theta,
\]
where $\scX \times \scZ$ is the original augmented space for involutive MCMC,
$\Delta := \lt\{a,b \in (0,1)^2 : a < b\rt\}$ is a set of acceptance ratio thresholds $(a,b)$,
and $\Theta$ is the set of tuning parameters $\theta$ for the involutions.
We assume $\Theta$ is a standard Borel space, such that $\scS$ is standard Borel as well.
Let $\sbf : \scS \to \scS$ denote the augmented involution $f_\theta$:
for a point $s=(x,z,a,b,\theta)\in\scS$, define
\[
\sbf(s) &= (f_\theta(x,z), a, b, \theta) & J(s) &= \lt|\grad f_\theta(x,z)\rt|.
\]
Note that $\sbf$ is itself an involution on $\scS$. 
We then define the augmented target density 
\[
  \bar\pi(s) = 2\pi(x)\cdot m(z) \cdot \1_\Delta(a,b) \cdot\eta(\theta\mid x,z,a,b),
\]
where we assume that $\pi(x)$ and $m(z)$ are with respect to the Lebesgue measure on $\scX\times \scZ$,
that both $m$ and $\eta$ admit \iid draws,
and that there exists a $\sigma$-finite measure $\d\theta$ on $\Theta$ such
that for all $x\in\scX$, $z\in\scZ$ and $(a,b)\in\Delta$, $\eta(\cdot \mid x,z,a,b)$ is a density
with respect to $\d\theta$, but otherwise may depend arbitrarily on $x,z,a,b$.
The $\scX$-marginal of $\bar\pi$ is the target of interest, $\pi$. 

\begin{algorithm}[t]
	\begin{algorithmic}[1]
    \Require Initial $x$ with $\pi(x)>0$, target $\pi$, auxiliary distribution $m$, 
    step size distribution $\eta$, involutions $\cbra{f_\theta}_{\theta \in \Theta}$
        \State $z \sim m$ \Comment{auxiliary refreshment}
        \State $(a, b) \sim \Unif(\Delta)$ \Comment{soft acceptance bounds}
        \State $\theta \sim \eta(\d\theta \mid x, z, a, b)$ \Comment{refresh tuning parameter}
	\State $s \gets (x,z,a,b,\theta)$ \Comment{augmented state}
        \State $s' \gets \sbf(s)$\Comment{involutive proposal}
        \State $U \sim \Unif[0,1]$
        \If{$U \leq \min \lt\{1, \, \frac{\sbpi(s')}{\sbpi(s)}J(s)\rt\}$}
            \State \Return  $x'$  \Comment{accept}
        \Else 
            \State \Return  $x$ \Comment{reject}
        \EndIf
	\end{algorithmic}
  \caption{One iteration of AutoStep MCMC}
  \label{alg:autoMCMC}
\end{algorithm}

Given this setup, starting from $x_t\in\scX$, AutoStep MCMC (\cref{alg:autoMCMC}) consists of the following
steps:
\begin{enumerate}
    \item \textbf{Auxiliary refreshment:} Draw
    \[ z_t \dist m \qquad \text{and}\qquad (a_t,b_t) \dist \Unif(\Delta).\]

    \item \textbf{Tuning parameter refreshment:} Draw 
    \[ \theta_t \sim \eta(\d\theta \mid x_t, z_t, a_t,b_t).\]

    \item \textbf{Proposal:} Set $s_t = (x_t,z_t,a_t,b_t,\theta_t)$ and 
     \[ 
  s'_{t+1} &= \sbf(s_t) = (x'_{t+1},z'_{t+1},a'_{t+1},b'_{t+1},\theta'_{t+1}).
\]

    \item \textbf{Accept:} Set $x_{t+1} = x'_{t+1}$ with probability
      \[ 
        \min \lt\{ 1, \frac{\sbpi(s'_{t+1})}{\sbpi(s_t)} J(s_t)\rt\},
      \]
      and otherwise set $x_{t+1} = x_t$.
\end{enumerate}

We can recover standard involutive MCMC by setting $\eta(\d\theta|x,z,a,b) = \delta_{\theta_0}(\d\theta)$
for some fixed $\theta_0\in\Theta$. Further, choosing different 
involution families $\{f_\theta : \theta\in\Theta\}$ and auxiliary distributions 
$m$ on $\scZ$ recovers variants of common algorithms, e.g., RWMH, MALA, and HMC.
The major improvement is that the chain may draw the tuning parameter $\theta\in\Theta$
automatically at each step in a manner that depends on the current state $(x_t, z_t,a_t,b_t)$.
The key design choice, then, is to select a conditional tuning refreshment distribution $\eta$
that yields values of $\theta$ that are well-adapted to the local shape of the target $\bar\pi$.
In this work, we focus on the design of the conditional tuning refreshment distribution $\eta$ 
in the case where $\theta$ is a step size parameter (with $\Theta = \reals_+$).
However, the AutoStep MCMC method described previously has the correct
stationary distribution for more general parameter spaces $\Theta$ (see \cref{prop:hard_invariance}).
Note also that the above approach of sampling a tuning parameter $\theta$ from a conditional distribution $\eta$ given the 
augmented state was also introduced in concurrent work on Gibbs self-tuning \citep{BouRabee24,BouRabee24NUTS},
although the design of $\eta$ in \cref{sec:stepsizeselection} differs.

\subsection{Step size selection} \label{sec:stepsizeselection}
We now focus on the design of the tuning refreshment distribution $\eta$ 
when $\theta$ is a step size parameter with $\Theta = \reals_+$. 
Intuitively, a good choice of $\theta$ should yield a proposal
for which the acceptance ratio $\exp(\ell(x,z,\theta))$ 
of the original involutive method is not too close to either
0 ($\theta$ is too large) or 1 ($\theta$ is too small).
Critically, this should also be true for $\exp(-\ell(x,z,\theta))$,
which is the acceptance ratio in the reverse direction 
\[
\ell(f_\theta(x,z),\theta) = -\ell(x,z,\theta).
\]
To avoid setting arbitrary fixed bounds on $\ell$, we
use random $a,b$ as thresholds 
and ensure that $|\ell|$
\emph{roughly} tries to fall in the range $(|\log(b)|, |\log(a)|)$.
More precisely, given a fixed initial step size $\theta_0 > 0$,
we propose setting the step size $\theta$ to
\[
  \theta &= \theta_0 \times 2^{\mu(x,z,a,b)}, \label{eq:implicit_eta}
\]
where $\mu$ is the step size selector function
\[
&\mu(x,z,a,b) =\\
&\lt\{\!\!\small\begin{array}{ll}
\min\{j\in\ints^+ : |\ell(x,z,\theta_02^j)| \geq |\log b| \}-1, & |\ell_0| < |\log b|\\
\max\{j\in\ints^- : |\ell(x,z,\theta_02^j)| \leq |\log a| \}, & |\ell_0| > |\log a|\\
0, & \text{otherwise},
\end{array}\rt.
\]
and $\ell_0 = \ell(x,z,\theta_0)$. Therefore $\eta(\d\theta | x,z,a, b)$ is the Dirac
measure at $\mu(x,z,a,b)$, which has a density with respect to the counting measure on 
$\{\theta_0\times 2^j : j\in\sdZ\}$.

The pseudocode for computing $\mu(x,z,a,b)$ is given in \cref{alg:stepsize_selector}. 
If the initial step size $\theta_0$ yields an acceptable $|\ell_0|$, the function simply returns $j=0$.
If the initial step size is too large ($|\ell_0| > |\log a|$), $j$ is decreased until
$|\ell_0| \leq |\log a|$. And if the initial
step size is too small ($|\ell_0| < |\log b|$), $j$ is increased until $|\ell_0| > |\log b|$, and then finally decreased
by $1$ to avoid poor proposals. Note that this function does 
not guarantee that $|\log b| \leq |\ell| \leq |\log a|$ precisely, but finds a good trade-off by
approximately targeting that range while avoiding the need for expensive methods to find values exactly within the bounds.

\begin{algorithm}[t]
	\begin{algorithmic}[1]
    \Require state $x, z, a, b$, initial step size $\theta_0$.
    \State $\theta \gets \theta_0$ 
    \State $\ell \gets \ell(x,z,\theta)$
    \State $v \gets \1\{|\ell| < |\log b|\} - \1\{|\ell| > |\log a|\}$ \label{line:v}
    \State $j = 0$ \Comment{number of doublings/halvings}
    \If{$v = 0$}
    	\State \Return $j$
    \EndIf
    \While{true} 
        \State $j \gets j + v$
        \State $\theta \gets \theta_0\cdot2^j$
        \State $\ell \gets \ell(x,z,\theta)$
        \If{$v=1$ and $|\ell| \geq |\log b|$}  
            \State \Return $j-1$ \Comment{Require final halving}  
        \ElsIf{$v=-1$ and $|\ell| \leq |\log a|$}
            \State \Return $j$  
        \EndIf 
    \EndWhile
	\end{algorithmic}
  \caption{Step size selector $\mu$}
  \label{alg:stepsize_selector}
\end{algorithm}

The step size refreshment was inspired by that of autoMALA \citep{BironLattes24},
but has two important differences. 
First, we use symmetric thresholds that check 
$|\log b| \leq |\ell| \leq |\log a|$, instead of 
checking $\log b \leq \ell \leq \log a$.
This is crucial for ensuring irreducibility of the method (see \cref{sec:theory}),
and avoids the sampler getting stuck in the tails or near the mode 
of the target (see \cref{fig:sym_vs_asym} for empirical results to this effect).
Second, we include the step size $\theta$ as an augmentation of our state variable,
which substantially simplifies theoretical analyses (compare the proof
of \cref{prop:hard_invariance} in \cref{sec:proofs} with the proof of Theorem 3.4 in \cite{BironLattes24}).

\subsection{Round-based tuning}
\label{sec:round-based}

The AutoStep MCMC method has one free parameter: the initial 
step size $\theta_0$.
While the method is insensitive to $\theta_0$ in that its 
performance per unit cost shrinks slowly at a rate of roughly $O(|\log \theta_0|)$ (see \cref{fig:fixvsauto_ksess,fig:fixvsauto_other} 
and \cref{cor:expectedtaurwmh} for empirical and theoretical evidence to that effect),
it is still helpful to tune
this parameter to minimize the number of doubling/halving steps.
Furthermore, many involutive MCMC methods---e.g., RWMH, MALA, and HMC---have a 
preconditioner, or \emph{mass matrix} $M$ that needs to be tuned.

In this work, we use a round-based procedure to tune $\theta_0$
and $M$ (Algorithm \ref{alg:round_autoMCMC}). 
Each round corresponds to a block of iterations during
which parameters are held constant. We use $\theta_0 = 1$
and $M = I$ for the first round. 
At the end of each round we update $\theta_0 \gets \theta_0\times 2^{\shmu}$,
where $\shmu$ is the empirical median of the selected log step sizes $\mu_t = \mu(x_t,z_t,a_t,b_t)$
from the current round. We also set $\shM$ to the diagonal of the inverse sample covariance
matrix, which is then mixed with the identity as a regularizer to form $M$
in each iteration (Line 8 in \cref{alg:round_autoMCMC}).

\begin{algorithm}[t]
	\begin{algorithmic}[1]
    \Require Initial $x_0$, number of rounds $R$, 
    target $\pi$, auxiliary distribution $m$, 
    step size distribution $\eta$, involutions $\cbra{f_\theta}_{\theta \in \Theta}$.
	\State $\theta_0 \gets 1$, $\shM \gets I$
	\State $m \gets \Norm(0, I_d)$
    \For{$r$ {\bf in} 1, 2, \dots, $R$}
      \State $T \gets 2^r$   
      \Comment{Number of iterations}
      \State $\eta \gets \mathrm{Dirac}(\theta_0\times 2^{\mu(x,z,a,b)})(\d \theta)$
      \For{$t$ {\bf in} 1, 2, \dots, $T$} 
        \State $\xi \sim \frac{1}{3} \delta_0 + \frac{1}{3} \delta_1 + \frac{1}{3} \text{Beta}(1,1)$
        \LineComment{Random mixing of the preconditioner}
        \State $M_{i,i}^{1/2} \gets \xi \shM_{i,i}^{1/2} + (1 - \xi)$
        \State $m \gets \Norm(0, M)$
        \LineComment{see definition of $\mu_t$ in \cref{sec:round-based}}
        \State $x_t, \mu_t \!\gets\! \texttt{AutoStep}(x_{t-1}, \pi, m, \eta, \cbra{f_\theta}_{\theta \in \Theta})$
      \EndFor
      \State $\theta_0 \gets \theta_0 \times 2^{\text{median}(\mu_1,\dots,\mu_T)}$
      \State $x_0 \gets x_T$ 
      \State $\shM \gets \text{diag}
        \lt( \lt[\Var[x_t^{(j)}]_{t=1}^T \rt]_{j= 1}^{d} \rt)^{-1}$
    \EndFor
    \State \Return $\{x_t\}_{t=1}^T$
	\end{algorithmic}
  \caption{Round-based AutoStep MCMC}
  \label{alg:round_autoMCMC}
\end{algorithm}

\section{Theoretical Analysis}
\label{sec:theory} 

The marginal sequence $x_t$ on $\scX$ of AutoStep MCMC is itself a Markov chain
because each step redraws $z_t, a_t, b_t, \theta_t$ 
independently of their previous value conditioned on $x_t$.
In this section we establish various properties of the $\scX$-marginal Markov chain.

\subsection{Invariance}
\label{sec:invariance}

First, we show that AutoStep MCMC is $\bar\pi$-invariant on the augmented
space $\scS$, and hence $\pi$-invariant on $\scX$. 
The result is a straightforward application of \citet[Theorem 2]{tierney1998}
on the augmented space $\scS$.
Note that while this work focuses on step size parameters $\theta\in\reals_+$,
\cref{prop:hard_invariance} below holds for general parameter spaces $\Theta$
and tuning refreshment distributions $\eta(\d\theta|x,z,a,b)$.

\bassum 
\label{assum:diff_involution}
For each $\theta \in \Theta$, $f_\theta$ is a 
continuously differentiable involution.
\eassum

\bprop
\label{prop:hard_invariance}
Under \cref{assum:diff_involution}, 
AutoStep MCMC is $\sbpi$-invariant,
and hence the $\scX$-marginal is $\pi$-invariant.
\eprop

\subsection{Irreducibility and aperiodicity} 
\label{sec:Irreducibility}
Next, we establish that the $\scX$-marginal of AutoStep MCMC 
is \emph{$\pi$-irreducible} and \emph{aperiodic} (see \citet{Roberts04}): intuitively,
the chain has a positive probability of 
eventually visiting any measurable $A\subseteq \scX$ with $\pi(A) > 0$,
and it does not visit various sets in a repeating pattern.
We will demonstrate $\pi$-irreducibility and aperiodicity simultaneously by showing that
the $\scX$-component of the chain can reach any measurable set 
$A \subset\scX$ in a \emph{single} step with positive probability (\emph{one-step irreducibility}).
The first assumption needed is that for any fixed  $\theta\in\Theta$, the $\scX$-marginal
kernel $P_\theta(x, \cdot)$ of the original involutive MCMC algorithm given by \cref{eq:basekernel1,eq:basekernel2}
can do so as well.
\bassum 
\label{assum:base_irreducible} 
For all $x\in\scX$, $\theta\in\Theta$, and $A\subseteq \scX$
such that $\pi(A) > 0$, 
the $\scX$-marginal kernel $P_\theta$ of 
involutive MCMC (\cref{eq:basekernel1,eq:basekernel2})
satisfies $P_\theta(x, A) > 0$. 
\eassum
The second assumption needed is that 
there is a non-null set of parameters $\theta\in\Theta$
that can be selected and result in an accepted 
move from any point $x,z\in\scX\times\scZ$ in the original
augmented space of involutive MCMC.
We encode this using the positivity of the function
\[
&\gamma(x,z,\theta)=\\
&\int \min\lt\{\eta(\theta \mid x,z,a,b), \eta(\theta\mid f_\theta(x,z),a,b)\rt\}\1_\Delta(\d(a,b)).
\]
\bassum\label{assum:paramsupport}
There exists a $B\subseteq \Theta$ such 
that $\int_B\d\theta > 0$ 
and for all $x\in\scX$, $m$-\aev $z\in\scZ$, and $\theta\in B$,
\[
\gamma(x,z,\theta) > 0.
\]
\eassum
These assumptions yield the desired result,
which holds for general parameter spaces $\Theta$ and distributions $\eta$.
\bprop\label{prop:onestepirr}
If both \cref{assum:base_irreducible,assum:paramsupport} hold,
then AutoStep MCMC is one-step irreducible, and hence irreducible and aperiodic.
\eprop
We now apply \cref{prop:onestepirr} to the case where $\theta$ is a step size parameter 
and we use $\eta$ from \cref{sec:stepsizeselection}.
In this setting, \cref{assum:paramsupport} simplifies substantially.

\bcor\label{cor:sigmairreduc}
Suppose $\Theta = (0, \infty)$, \cref{assum:base_irreducible} holds, and we use $\eta$ from \cref{sec:stepsizeselection}.
Then, AutoStep MCMC is irreducible and aperiodic if 
for all $x \in \scX$ and $m$-\aev $z \in \scZ$, 
$\abs{\ell(x,z,\theta_0)} \notin \cbra{0,\infty}$.
\ecor

We show in \cref{lem:RWMH_irreducible,lem:MALA_irreducible} that \cref{assum:base_irreducible}
holds for both RWMH and MALA under weak conditions, and hence
the irreducibility and aperiodicity of AutoStep RWMH and MALA
follows from 
 $\abs{\ell(x,z,\theta_0)} \notin \cbra{0,\infty}$.

\subsection{Step size selector termination}
\label{sec:termination}
We now establish that under mild conditions,
the step size selector function $\mu$ can be computed in finite time.
For starting state $s = (x,z,a,b)$ and initial step size $\theta_0 > 0$, let 
$\tau(s, \theta_0) \geq 1$ be the number of iterations of the while loop in \cref{alg:stepsize_selector}.
The key condition, \cref{assum:elllimits}, is satisfied intuitively when
the density $\pi\cdot m$ is continuous and the 
involution $f_\theta$ becomes the identity as $\theta\to 0$ and grows without bound as $\theta\to\infty$.
\bassum
\label{assum:elllimits}
For $\pi\times m$-\aev $(x,z) \in \scX \times \scZ$, 
\[
\lim_{\theta\to 0^+} |\ell(x,z,\theta)| &= 0 & \lim_{\theta\to\infty} |\ell(x,z,\theta)| = \infty.
\]
\eassum

\bprop \label{prop:stepsizeterminates}
Let $\theta_0 > 0$ and suppose 
\cref{assum:elllimits} holds. 
Then $\tau(s, \theta_0) < \infty$, $\bar \pi$-\as
\eprop

Note that while \cref{prop:stepsizeterminates} guarantees that the Markov
chain can be simulated in finite time in practice, the long-run computational cost of the step size adaptation
depends on the expected number of doubling/halving iterations in each step, $\E \tau(s,\theta_0)$ for $s\dist\sbpi$.
\cref{prop:expectedstepsizetermination} bounds this expectation in terms of $\ell(x,z,2^t\theta_0)$ for $t\in\sdZ$.

\bprop\label{prop:expectedstepsizetermination}
For $s=(x,z,a,b,\theta)\dist \sbpi$ and all $\theta_0 > 0$,
\[
\E\tau(s, \theta_0) \!\leq \!
\E \sum_{t=0}^\infty e^{-2|\ell(x,z,2^t\theta_0)|} \!+\! \lt(1\!-\!e^{-|\ell(x,z,2^{-t}\theta_0)|}\rt)^2\!.
\]
\eprop

The first and second terms in the sum in \cref{prop:expectedstepsizetermination} 
capture the expected number of doubling and halving steps, respectively.
While more detailed bounds on $\E\tau(s,\theta_0)$ require problem-specific analysis that depends on
$f_\theta$, $\pi$, and $m$, both terms will typically decay quickly in $t$.
\cref{cor:expectedtaurwmh} provides an example of a more detailed result based on \cref{prop:expectedstepsizetermination}
in a representative setting when using AutoStep with random walk Metropolis--Hastings.

\bcor\label{cor:expectedtaurwmh}
Let $\log\pi(x)$ be strongly concave with Lipschitz gradients,
and set $f_\theta$ as in \cref{eq:rwmh} with $M=I$.
Then
\[
\E\tau(s,\theta_0) = O\lt(\lt|\log \theta_0\rt|\rt)\quad \text{as}\quad \theta_0\to0\quad\text{or}\quad\theta_0 \to \infty,
\]
with a dimension-independent leading constant.
\ecor

\cref{cor:expectedtaurwmh} shows that the expected number of doublings/halvings---i.e., the 
long-run average evaluations of $f_\theta$ and $\ell$ per iteration---is very robust 
to the initial step size parameter $\theta_0$. This translates to a robust performance per unit cost: in the setting in \cref{cor:expectedtaurwmh},
we expect AutoStep RWMH to exhibit an average jump distance per unit cost to scale like $O\lt(|\log\theta_0|^{-1}\rt)$ for small/large $\theta_0$.
Contrast this to the significantly worse $O\lt(\exp(-|\log\theta_0|)\rt)$ decay for fixed step size RWMH.
This difference in behaviour is confirmed empirically in \cref{fig:fixvsauto_other}.

\subsection{Energy jump distance}
\label{sec:energyjumpdist}
For $s = (x,z,a,b,\theta) \dist \sbpi$,
$s' = \sbf(s) = (x',z',a',b',\theta')$,
and $U\dist\Unif[0,1]$,
define the \emph{energy jump distance} 
\[
D = |\ell(x,z,\theta)| \1\lt[U \leq \frac{\sbpi(s')}{\sbpi(s)}J(s)\rt],
\]
which we use to encode the change in $\log(\pi(x)m(z))$ after 
one iteration of AutoStep MCMC.
\cref{prop:energyjumpdist} shows that \emph{any}
involutive MCMC method---both traditional and AutoStep
methods---have an expected energy jump distance
bounded above by a simple expression
via the tuning parameter proposal density ratio $\sbeta$,
\[
\sbeta = \esssup_{x,z,a,b,\theta} \frac{\eta(\theta\mid f_{\theta}(x,z),a,b)}{\eta(\theta\mid x,z,a,b)} \quad\text{(under $\sbpi$)}.
\]
\bprop
\label{prop:energyjumpdist} 
Under \cref{assum:diff_involution},
\[\E D \leq 2\sbeta\max\lt\{e^{-1}, \sbeta\log\sbeta \rt\}.\]
\eprop
In particular, for traditional involutive MCMC with a fixed parameter $\theta=\theta_0$,
or for AutoStep MCMC with $\sigma^2 = 0$, or AutoStep MCMC with a 
fixed-width uniform distribution for $\eta$, we have that $\sbeta \leq 1$, so
\[
\E D \leq 2e^{-1} \approx 0.736.
\]
The step size selector presented in \cref{sec:stepsizeselection}
is a computationally efficient method to make $|\ell(x,z,\theta)|$ fall roughly in the range 
$(|\log b|, |\log a|)$, where $a, b \dist \Unif(\Delta)$.
Therefore, as a heuristic, we expect
\[
0.5 = \E|\log b|
\lesssim \E |\ell(x,z,\theta)| \lesssim \E|\log a| = 1.5,
\]
with departures from exactness arising due to the discrete doubling/halving procedure
(as opposed to an exact root finder).
In other words, the step size selector in this work
creates proposals with mean energy jump distance
roughly targeting the maximum $\approx 0.736$
for a broad class of involutive MCMC methods.

\section{Experiments}
\label{sec:experiments}

In this section we present an empirical evaluation of two AutoStep MCMC variants: RWMH and MALA.
We first use synthetic targets with varying tail behaviour 
to examine the effect of the symmetric termination criterion in \cref{alg:stepsize_selector} versus
the asymmetric criterion from \citet{BironLattes24}, the efficacy of our proposed tuning
procedure for $\theta_0$, and the robustness of the performance of our proposed method versus the initial step
size $\theta_0$. We then investigate the performance of AutoStep RWMH and AutoStep MALA 
in comparison to previous adaptive methods.

Throughout, we measure the efficiency of each sampler in terms of 
effective sample size (ESS) \citep{Flegal08} per second. 
However, we found empirically that standard ESS estimates 
\citep[p.~286-287]{Gelman13}, \citep[p.~1539-1541]{Jones06}
did not accurately characterize sampler performance because they do not incorporate 
knowledge of the target distribution directly.
Therefore we instead use a diagnostic (\texttt{KSESS}) outlined in \cref{sec:ksess}
that involves the maximum difference between the empirical CDF and target CDF
across all dimensions, where the target CDF is approximated using a long run of
parallel tempering with \texttt{Pigeons.jl} \cite{Pigeons}. 
\texttt{KSESS} is used here solely for research comparisons and is not recommended 
for practical data analysis, as it relies on knowledge of the target or expensive 
gold-standard samples. The code for reproducing the main experimental results is 
available at \url{https://github.com/Julia-Tempering/AutoStep}.

\subsection{Symmetric step size selection criterion}\label{sec:expt_symvsasym}

\begin{figure}[t]
\centering \includegraphics[width=0.75\columnwidth]{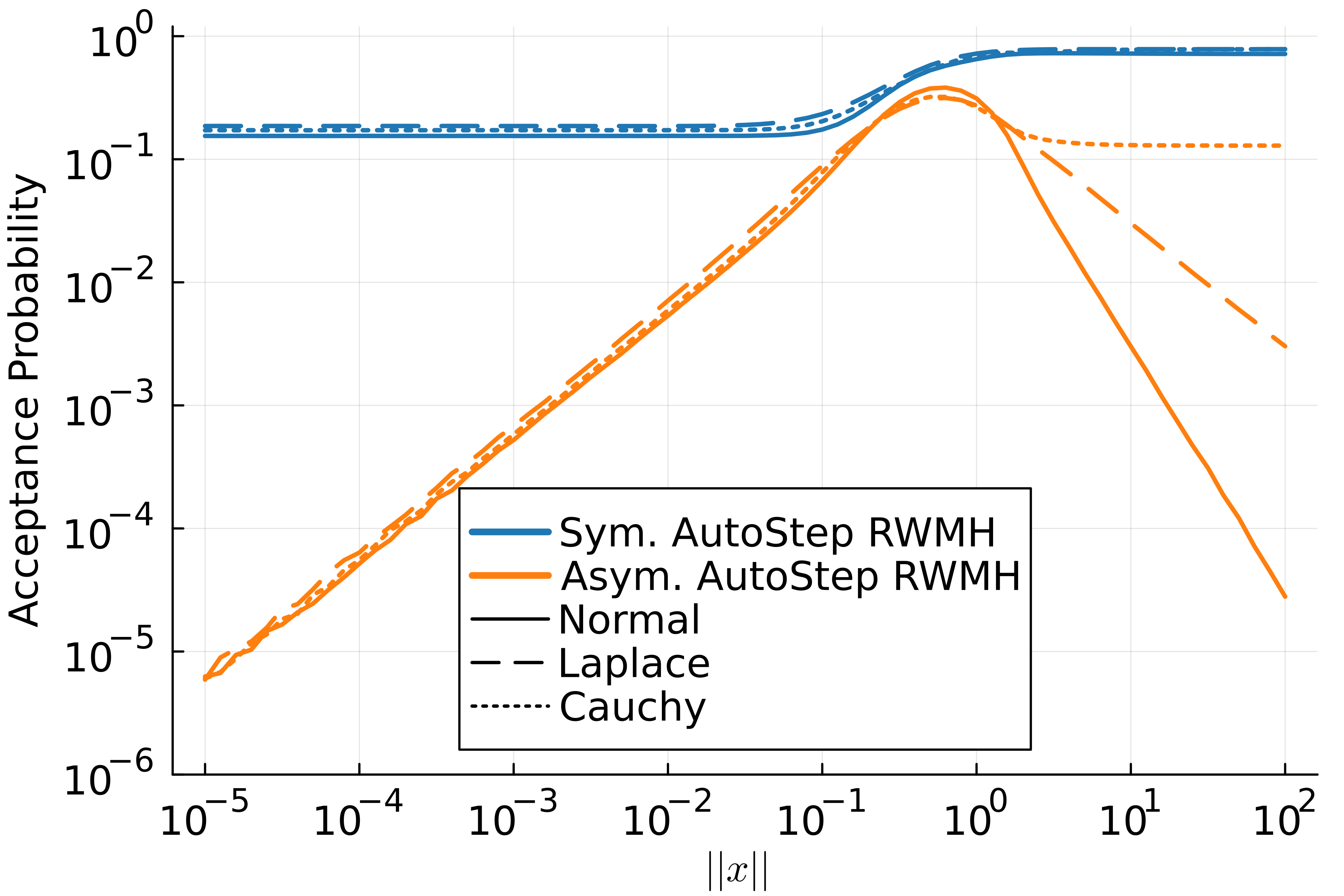}
\caption{Comparison of the symmetric (this work, blue) versus asymmetric (\citep{BironLattes24}, orange) 
step size criteria in \cref{alg:stepsize_selector}, in terms of the move
acceptance probability of AutoStep RWMH as a function of the state norm
$\|x\|$.  Note that the asymmetric criterion yields very low acceptance
probabilities for states near the mode (left side of the plot) and in the tails
(right side of the plot).
}\label{fig:sym_vs_asym}
\end{figure}

We first assess the benefit of using the symmetric step size 
criterion in \cref{alg:stepsize_selector} versus the asymmetric criterion
from \citet{BironLattes24} on three synthetic targets with
varying tail behavior: $\Norm(0,1)$, $\Laplace(0,1)$, and $\Cauchy(0,1)$. For each
$\|x\| \in \{10^{-5}, 10^{-4}, \dots, 10^2\}$, we simulated $10^7$ draws from
the target and renormalized them to the specified value. Then, for each draw, we
simulated one step of AutoStep RWMH/MALA ($\theta_0 = 1$), and
recorded the acceptance probability. 

The average of these acceptance probabilities is shown in
\cref{fig:sym_vs_asym} for AutoStep RWMH, showing that the
asymmetric step size selector can get stuck for extended periods near
the mode of the target (small $\|x\|$) and in the tails (large $\|x\|$).
This behavior particularly problematic in practice
for Bayesian inference, which is often initialized in the tails.
In contrast, the proposed symmetric step size selector is more robust, exhibiting an acceptance probability
that is greater than $10\%$ for the entire range of norms from $10^{-5}$ to $10^2$.
This result aligns with our theoretical results (\cref{prop:onestepirr,cor:sigmairreduc}) 
on the irreducibility of AutoStep with the symmetric step size selection criterion.

\subsection{Tuning the initial step size $\theta_0$}

\begin{figure}[t]
\begin{subfigure}{\linewidth}
   \centering \includegraphics[width=0.75\columnwidth]{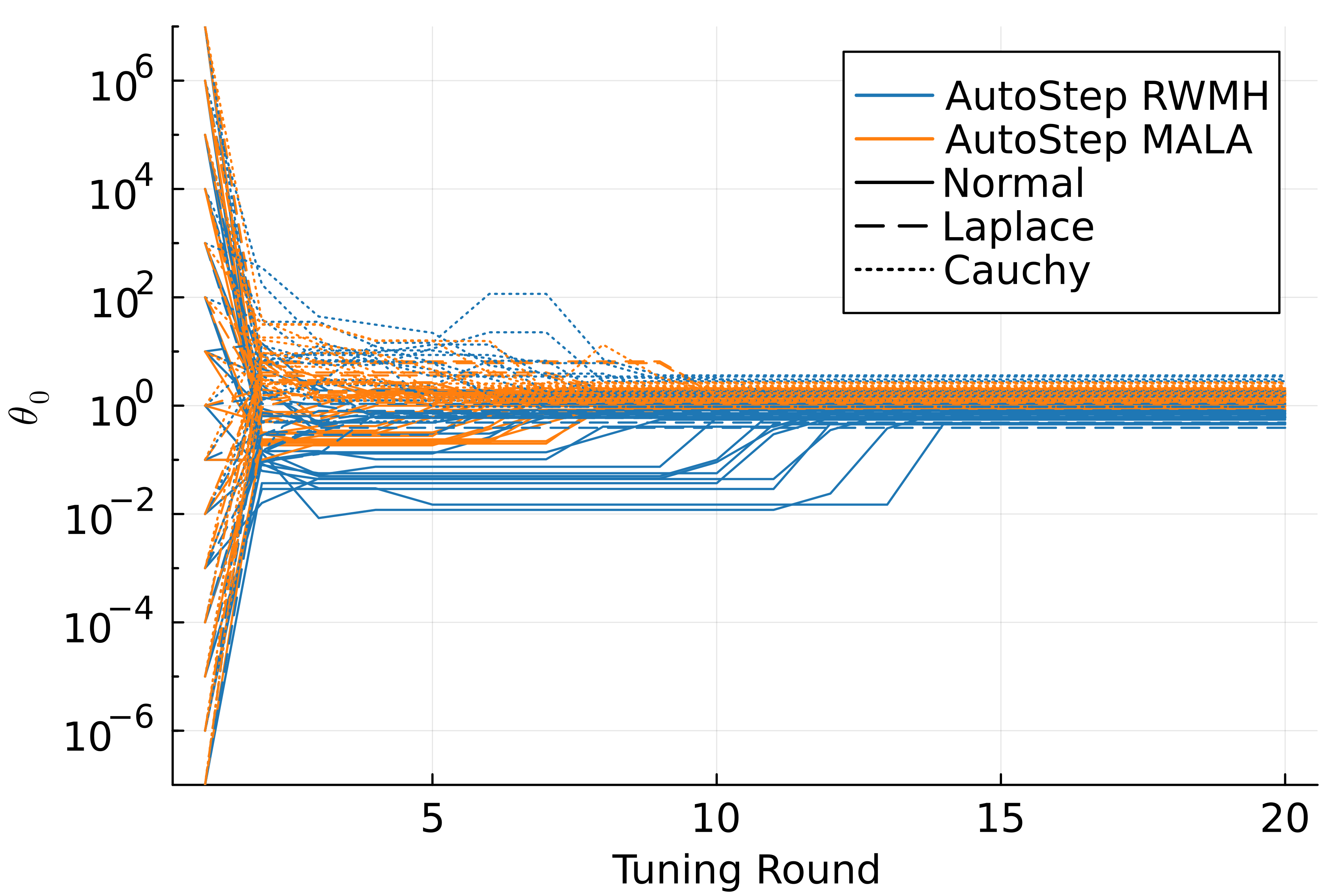}
   \caption{Values of $\theta_0$}\label{fig:stability}
\end{subfigure}
\begin{subfigure}{\linewidth}
   \centering \includegraphics[width=0.75\columnwidth]{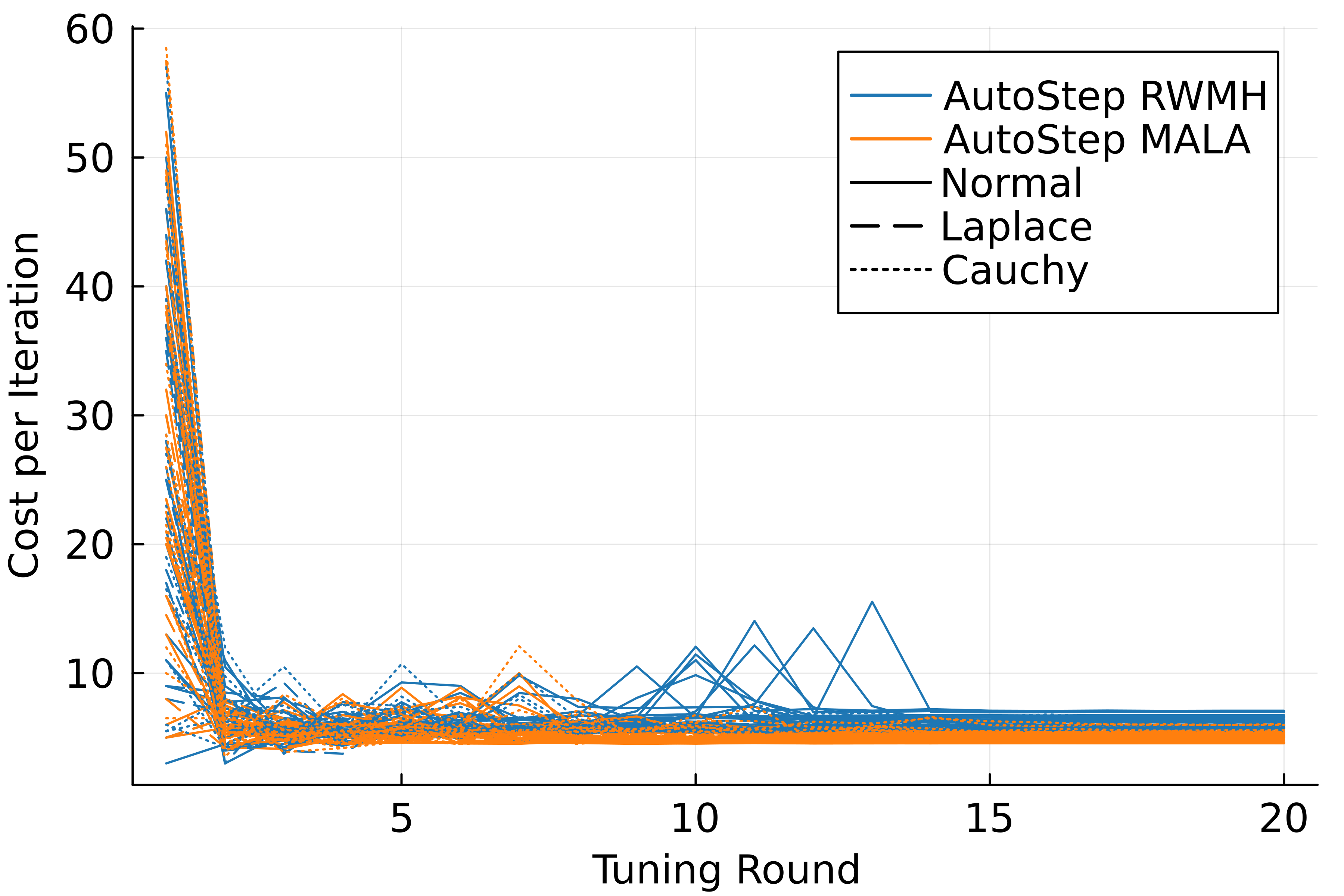}
   \caption{Cost per iteration}\label{fig:stabilitycost}
\end{subfigure}
\caption{The effect of tuning $\theta_0$,
showing traces of $\theta_0$ (\cref{fig:stability}) and 
cost per iteration (\cref{fig:stabilitycost}) versus tuning round when
$\theta_0$ is initialized in $\{10^{-7}, \dots, 10^7\}$. Despite
the wide range of initializations, the tuned values and cost per iteration
stabilize quickly and reliably.}\label{fig:theta0tuning}
\end{figure}

Next, we assess the stability and efficacy of the round-based
tuning procedure for $\theta_0$ in AutoStep RWMH/MALA.  For each
of the same three synthetic targets as in \cref{sec:expt_symvsasym}, we
initialize the state $x \dist \Norm(0, 20^2)$, $\theta_0\in \{10^{-7}, \dots,
10^{7}\}$, and run the Markov chain for $R=20$ doubling rounds ($\approx
2\times 10^6$ steps total), tuning $\theta_0$ per \cref{alg:round_autoMCMC}
after each round. For each trace, we track the value of $\theta_0$ as it is tuned,
as well as the per iteration cost, measured by the number of evaluations of $\ell$ 
in each round. Note that fixed step size MCMC methods call $\ell$ once per 
iteration, whereas AutoStep calls $\ell$ once plus an additional time for each 
doubling or halving during step size tuning.

\cref{fig:theta0tuning} displays the results from this experiment:
\cref{fig:stability} shows the tuned values of $\theta_0$ as a
function of doubling round, while \cref{fig:stabilitycost} shows the
corresponding average cost per iteration during each round. Both figures together indicate that the proposed
tuning procedure is highly effective: despite the initialization of $\theta_0$
spanning 14 orders of magnitude, the tuning remains stable and 
converges to a reasonable value of $\theta_0 \approx 1$, while the cost per iteration
is quickly minimized and remains stable throughout the rounds.

\begin{figure*}[t]
\centering
\begin{subfigure}{0.32\textwidth}
    \centering \includegraphics[width=\columnwidth]{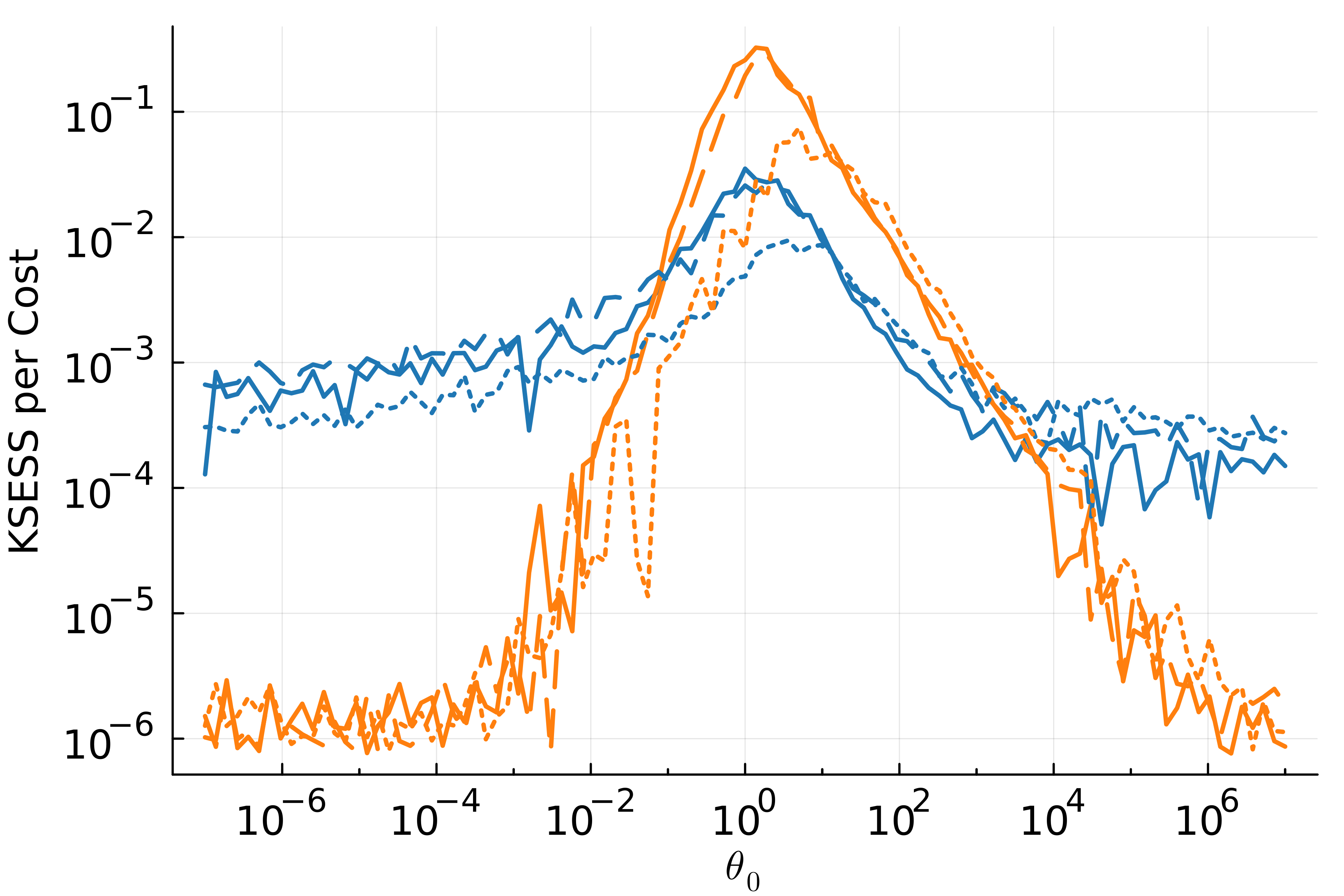}
   \centering\includegraphics[width=1.1\columnwidth]{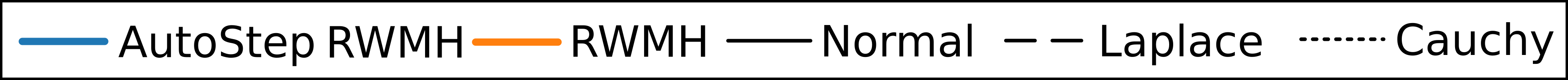}
    \caption{RWMH}\label{fig:ksess_rwmh}
\end{subfigure}
\hspace{1cm}
\begin{subfigure}{0.32\textwidth}
    \centering\includegraphics[width=\columnwidth]{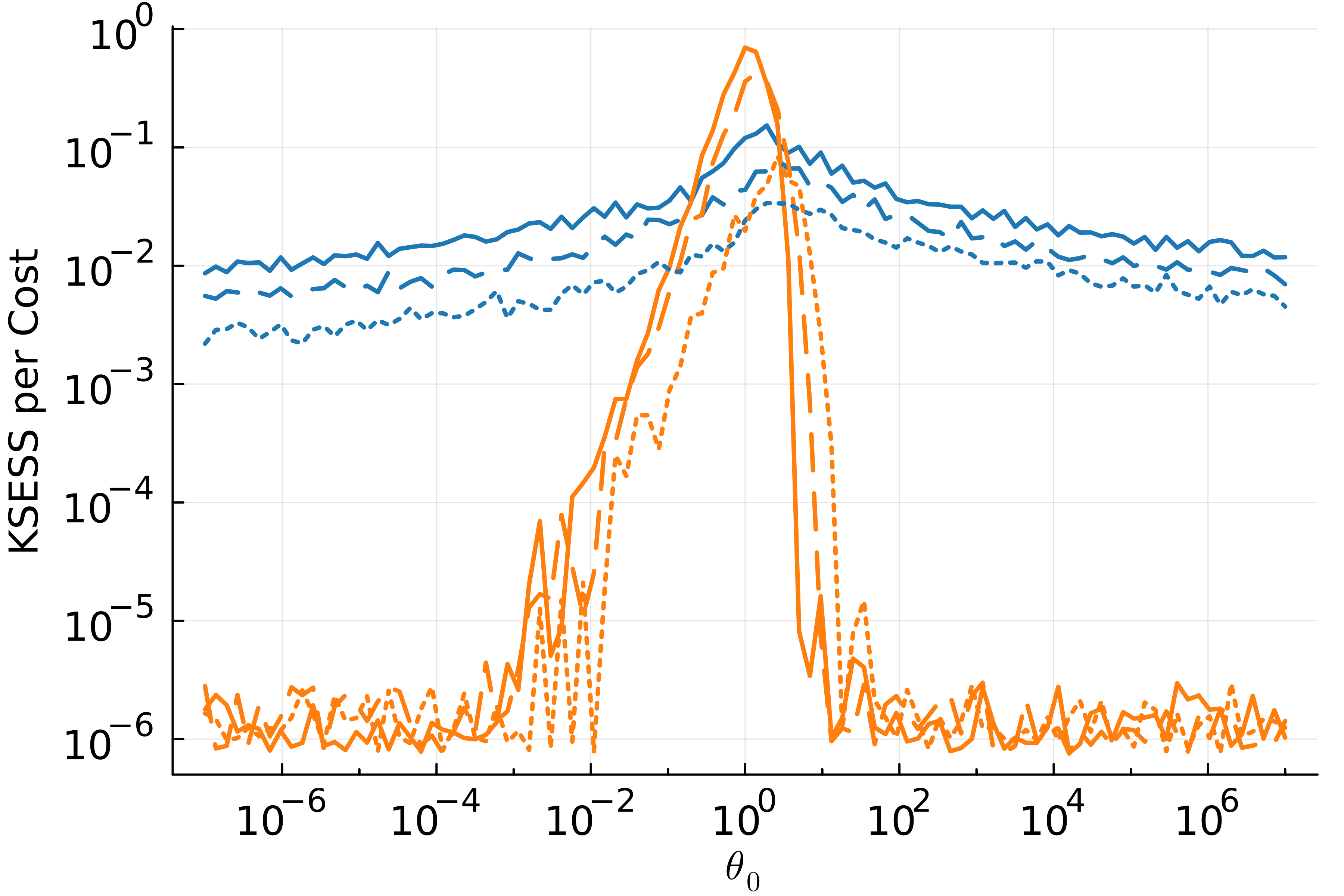}
   \centering\includegraphics[width=1.1\columnwidth]{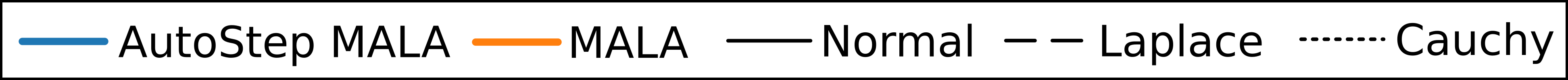}
    \caption{MALA}\label{fig:ksess_mala}
\end{subfigure}
\caption{\texttt{KSESS} per unit cost for AutoStep (blue) and fixed-step (orange) RWMH (\cref{fig:ksess_rwmh}) and MALA (\cref{fig:ksess_mala}). }\label{fig:fixvsauto_ksess}
\end{figure*}

\begin{figure*}[t]
\begin{subfigure}{0.32\textwidth}
    \centering \includegraphics[width=\columnwidth]{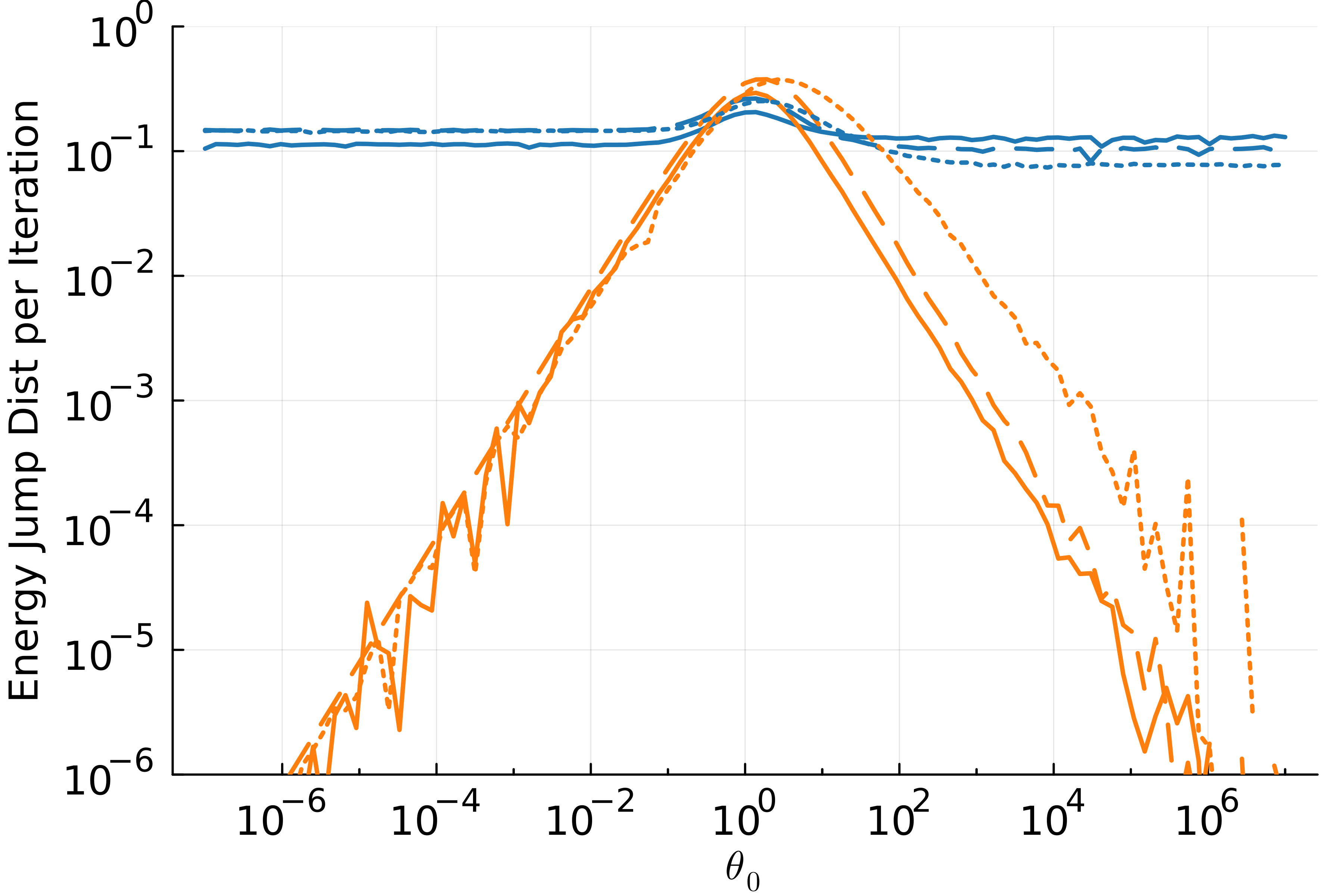}
   \centering\includegraphics[width=\columnwidth]{deliverables/ksess_and_ejump/legend_horizontal_RWMH.png}
    \caption{Energy Jump Distance per Iteration}\label{fig:ejump_rwmh}
\end{subfigure}
\hfill
\begin{subfigure}{0.32\textwidth}
    \centering\includegraphics[width=\columnwidth]{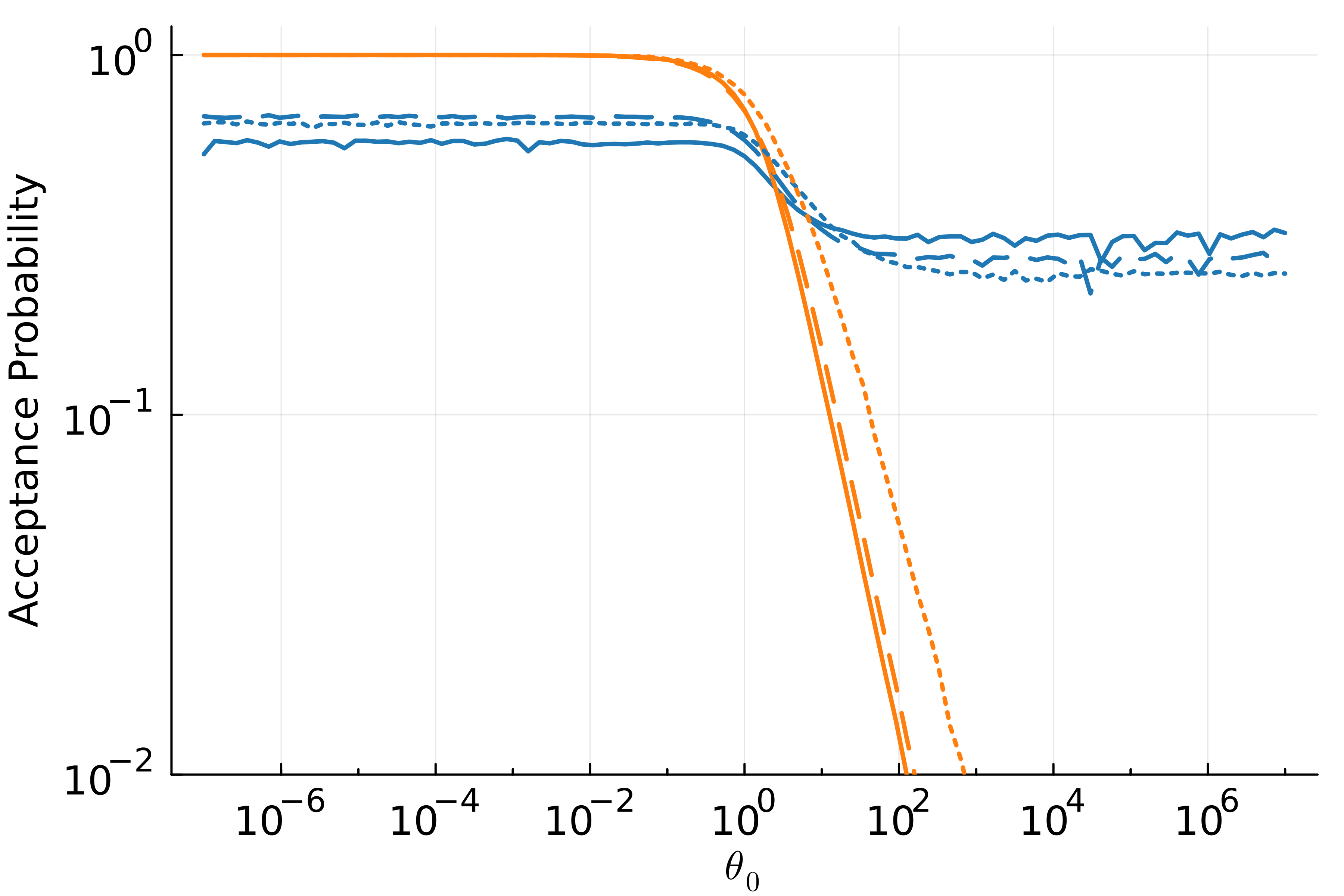}
   \centering\includegraphics[width=\columnwidth]{deliverables/ksess_and_ejump/legend_horizontal_RWMH.png}
    \caption{Acceptance Probability}\label{fig:acc_rwmh}
\end{subfigure}
\hfill
\begin{subfigure}{0.32\textwidth}
    \centering\includegraphics[width=\columnwidth]{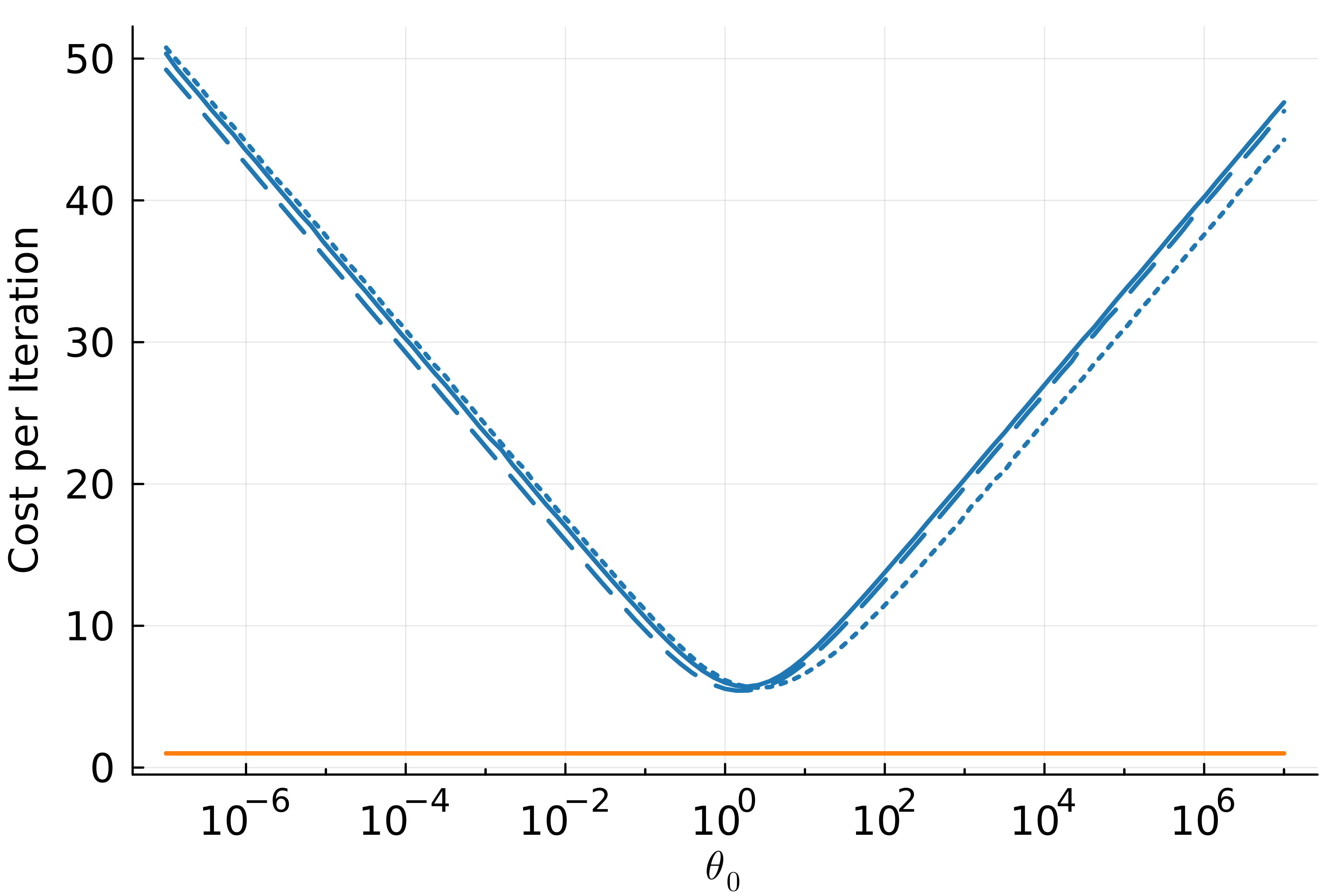}
   \centering\includegraphics[width=\columnwidth]{deliverables/ksess_and_ejump/legend_horizontal_RWMH.png}
    \caption{Cost per Iteration}\label{fig:cost_rwmh}
\end{subfigure}
\caption{Energy jump distance per iteration (\cref{fig:ejump_rwmh}),
acceptance probability (\cref{fig:acc_rwmh}),
and cost per iteration (\cref{fig:cost_rwmh}) for AutoStep and fixed step RWMH versus initial step size $\theta_0$.}\label{fig:fixvsauto_other}
\end{figure*}

\subsection{Robustness to initial step size $\theta_0$}\label{sec:expt_robustnesstheta0}

Next we examine the robustness of AutoStep RWMH/MALA to the 
setting of $\theta_0$ compared with fixed-step-size RWMH/MALA.
In this experiment we do not tune $\theta_0$, and fix it at values $\theta_0 \in \{10^{-7}, \dots, 10^7\}$.
For each of the three targets from \cref{sec:expt_symvsasym},
and each fixed $\theta_0$, we initialized each sampler
at a target draw, and collected various metrics over $10^6$ MCMC steps.

The results are presented in \cref{fig:fixvsauto_ksess,fig:fixvsauto_other}.
The main comparison is in \cref{fig:fixvsauto_ksess}, which illustrates the difference in \texttt{KSESS} per 
unit cost for fixed step size vs.~AutoStep methods, where one unit cost corresponds to 
one evaluation of $\ell$. For these 1D unimodal targets, 
standard fixed step size methods perform well when well-tuned. But for poor step size choices,
performance decays quickly at a rate of roughly $O(\exp(-|\log\theta_0|))$. In contrast, AutoStep methods 
incur a penalty for adaptivity, but the performance is far more robust to $\theta_0$ and decays like $O(|\log\theta_0|^{-1})$, which aligns
with our theory from \cref{cor:expectedtaurwmh}.
This main comparison is supported by additional results in \cref{fig:fixvsauto_other}.
\cref{fig:ejump_rwmh} demonstrates that AutoStep methods empirically provide a high energy jump distance per iteration
across all values of $\theta_0$,
which is valuable especially in the context of annealing/tempering methods that depend on the mixing of the energy statistic \citep{Surjanovic24}.
\cref{fig:acc_rwmh} shows that the acceptance probability of AutoStep robustly remains bounded away from 0 and 1 across all values of $\theta_0$.
Finally \cref{fig:cost_rwmh} confirms the $O(|\log\theta_0|)$ scaling of cost per iteration predicted by \cref{cor:expectedtaurwmh}.

\begin{figure}[h]
   \centering
   \includegraphics[width=\columnwidth]{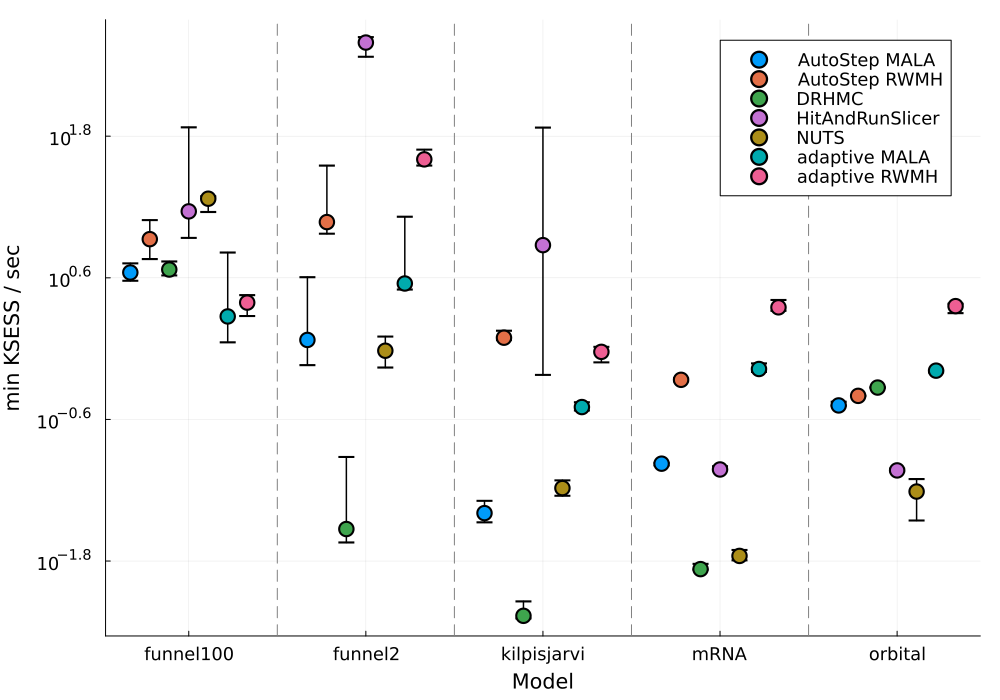}
   \caption{\texttt{min KSESS} per second for AutoStep and state-of-the-art 
   samplers across five benchmarked models.}
   \label{fig:minKSess_per_sec}
\end{figure}

\subsection{Comparison with other adaptive methods}
\label{sec:experiment_compare_existing}

We compare the performance of AutoStep RWMH/MALA against five 
state-of-the-art adaptive samplers: NUTS \citep{Neal11}, the hit-and-run 
slice sampler \citep{Neal03,Belisle93}, adaptive RWMH 
\citep{vihola12}, adaptive MALA \citep{advancedhmc}, 
and delayed rejection HMC \citep{Modi24}. For AutoStep RWMH/MALA, 
we use round-based adaptive tuning to adjust the initial step size 
and diagonal preconditioner. Our benchmarks include two synthetic distributions—Neal’s 
funnel in 2 and 100 dimensions—and three real Bayesian posteriors: a 
three-parameter linear regression model for yearly temperatures at 
Kilpisj\"{a}rvi, Finland \citep{kilpisjarvi}, an orbit-fitting problem 
\citep{Octofitter}, and an mRNA transfection model \citep{Ballnus17}.

\cref{fig:minKSess_per_sec} presents the results in terms of 
minimum \texttt{KSESS} across dimensions per second.
This figure demonstrates that AutoStep methods consistently achieve reasonable 
performance across all five benchmarks, particularly on the challenging 
funnel targets. AutoStep RWMH shows competitive performance, generally 
within the range of the other methods. AutoStep MALA shows slightly lower performance 
in some settings but exhibits better scaling to higher dimensions, as seen in the 
\texttt{funnel100} results. Overall, both AutoStep variants demonstrate robust 
efficiency across diverse model geometries.

\section{Conclusion} 
\label{sec:conclusion}

In this paper we presented AutoStep MCMC, a locally-adaptive step size selection
method for involutive MCMC. We proved that AutoStep MCMC kernels are $\pi$-invariant,
irreducible, and aperiodic under mild conditions. We also provided bounds
on the mean energy jump distance and expected cost per iteration. 
We demonstrated empirically that AutoStep MCMC is stable, reliable, and competitive with other adaptive 
methods.
One promising direction for future work is a rigorous
theoretical analysis of the sampling efficiency of AutoStep MCMC in terms
of expected squared jump distance and asymptotic variance.

\section*{Acknowledgements}
ABC and TC acknowledge the support of an NSERC Discovery Grant. TL acknowledges the support of the UBC Advanced Machine Learning Training Network. NS acknowledges the support of a Vanier Canada Graduate Scholarship. We additionally acknowledge use of the ARC Sockeye computing platform from the University of British Columbia.

\section*{Impact Statement}

This work presents a new MCMC algorithm;
the societal consequences of MCMC need not be discussed in this paper.

\bibliography{main}
\bibliographystyle{icml2025}

\newpage
\appendix
\onecolumn

\section{Proofs} 
\label{sec:proofs} 

\bprfof{\cref{prop:hard_invariance}}
The auxiliary refreshment and tuning parameter refreshment steps in AutoStep MCMC (Steps 1.~and 2.) resample
$(z, a, b, \theta)$ jointly from their conditional distribution given $x$ under $\sbpi$.
This move is well-known to be $\bar\pi$-invariant, and so it remains only
to show that the Metropolis-corrected involutive proposal 
(Steps 3.~and 4.) is $\bar\pi$-invariant. 
The kernel for the proposal on the augmented space $\scS$ is
\[
Q(s, \d s') = \delta_{\sbf(s)}(\d s'),
\]
and the acceptance probability $\alpha : \scS^2 \to \reals_+$ is given by
\[
\alpha(s, s') = \min\lt\{1, \frac{\sbpi(s')}{\sbpi(s)} J(s)\rt\}.
\]
In the notation of \citet[Theorem 2]{tierney1998},
define the measure $\mu(\d s, \d s') = \sbpi(\d s) \delta_{\sbf(s)}(\d s')$; 
because $\sbf$ is an involution, we have that the symmetric set $R$ 
and density ratio $r : R\to\reals_+$ are given by
\[
R = \{(s, s') \in \scS^2 : \sbf(s) = s', \,\sbpi(s) > 0,\, \sbpi(s') > 0\},
\qquad r(s,s') = \frac{\sbpi(\d s) \delta_{\sbf(s)}(\d s')}{\sbpi(\d s')\delta_{\sbf(s')}(\d s)}.
\]
Note that condition (i) of \citet[Theorem 2]{tierney1998} holds by definition of $R$ and $\alpha$.
Suppose for the moment that $r(s,s') = \frac{\sbpi(s)}{\sbpi(s')} J(s')$; then condition (ii)---and hence $\sbpi$-invariance---holds because
\[
\alpha(s,s')r(s,s') &= \min\lt\{1, \frac{\sbpi(s')}{\sbpi(s)}J(s)\rt\}\frac{\sbpi(s)}{\sbpi(s')} J(s') 
=\min\lt\{\frac{\sbpi(s)}{\sbpi(s')} J(s'), J(s')J(s)\rt\}
=\alpha(s',s),
\]
which follows because $J(s)J(s') = 1$ $\mu$-\aev on $R$. 
To demonstrate that $r(s,s')$ has the required form, consider a test function $g:\scS^2 \to \reals$:
\[
\int g(s,s') \sbpi(\d s) \delta_{\sbf(s)}(\d s') &= \int g(s, \sbf(s))\sbpi(\d s)\\
&= \int g((x,z,a,b,\theta), (f_\theta(x,z),a,b,\theta)) \sbpi(x,z,a,b,\theta) \d x \d z \d(a,b) \d\theta.
\]
Since $f_\theta$ is a continuously differentiable involution, 
we can transform variables $x', z' = f_\theta(x,z)$ by including a Jacobian term $J(s) = \lt|\nabla f_\theta(x,z)\rt|$ in the integrand
and by noting $x,z = f_\theta(x',z')$:
\[
&= \int g((f_\theta(x',z'),a,b,\theta), (x',z',a,b,\theta)) \sbpi(f_\theta(x',z'),a,b,\theta) \lt|\nabla f_\theta(x',z')\rt| \d x' \d z' \d(a,b) \d\theta\\
&= \int g(\sbf(s'), s') \sbpi(\sbf(s')) J(s') \d s'\\
&= \int g(\sbf(s'), s') \frac{\sbpi(\sbf(s'))}{\sbpi(s')} J(s') \sbpi(\d s')\\
&= \int g(s, s') \frac{\sbpi(s)}{\sbpi(s')} J(s') \sbpi(\d s') \delta_{\sbf(s')}(\d s).
\]
Examining the first and last integral expressions, the density ratio has the form
\[
r(s,s') = \frac{\sbpi(\d s) \delta_{\sbf(s)}(\d s')}{\sbpi(\d s')\delta_{\sbf(s')}(\d s)} = \frac{\sbpi(s)}{\sbpi(s')} J(s').
\]
\eprfof

\bprfof{\cref{prop:onestepirr}} 
Let $K(x, \cdot)$ denote the Markov kernel for the $\scX$-marginal process of AutoStep MCMC.
Since for $u,v \geq 0$,  $\min\{1, uv\} \geq \min\{1, u\}\min\{1, v\}$, we have that for $s = (x,z,a,b,\theta)$,
\[
\min\lt\{1, \frac{\sbpi(\sbf(s))}{\sbpi(s)}J(s)\rt\} \geq \min\lt\{1, e^{\ell(x,z,\theta)}\rt\}\min\lt\{1, \frac{\eta(\theta\mid f_\theta(x,z),a,b)}{\eta(\theta\mid x,z,a,b)}\rt\}.
\]
Therefore
\[
K(x, A) &\geq \int \1[f_\theta(x,z)\in A\times \scZ] \min\lt\{1, e^{\ell(x,z,\theta)}\rt\}\min\lt\{1, \frac{\eta(\theta\mid f_\theta(x,z),a,b)}{\eta(\theta \mid x,z,a,b)}\rt\}\eta(\d\theta \mid x,z,a,b)m(\d z)\1_\Delta(\d(a,b)) \\
&= \int \1[f_\theta(x,z)\in A\times \scZ] \min\lt\{1, e^{\ell(x,z,\theta)}\rt\}\gamma(x,z,\theta)m(\d z) \d\theta,\label{eq:kbound}
\]
where
\[
\gamma(x,z,\theta) = \int \min\lt\{\eta(\theta \mid x,z,a,b), \eta(\theta\mid f_\theta(x,z),a,b)\rt\}\1_\Delta(\d(a,b)).
\]
By \cref{assum:paramsupport}, for all $x\in\scX$, $m$-\aev $z\in\scZ$, and for all $\theta \in B$ where $\int_B \d\theta > 0$,
$\gamma(x,z,\theta) > 0$. Therefore
\[
&\int \1[f_\theta(x,z)\in A\times \scZ] \min\lt\{1, e^{\ell(x,z,\theta)}\rt\}\gamma(x,z,\theta)m(\d z)\1[\theta\in B] \d\theta > 0\\
\iff &\int \1[f_\theta(x,z)\in A\times \scZ] \min\lt\{1, e^{\ell(x,z,\theta)}\rt\}m(\d z)\1[\theta\in B] \d\theta > 0.
\]
The proof concludes by noting that
\[
\int \1[f_\theta(x,z)\in A\times \scZ] \min\lt\{1, e^{\ell(x,z,\theta)}\rt\}m(\d z)\1[\theta\in B] \d\theta 
&= \int P_\theta(x, A)\1[\theta\in B] \d\theta, 
\]
where $P_\theta(x, A)$ is the one-step probability 
of transitioning into $A$ from $x$ for the original involutive chain with parameter $\theta$,
and then by applying \cref{assum:base_irreducible}.
\eprfof

\bprfof{\cref{cor:sigmairreduc}}
The proof involves verifying \cref{assum:paramsupport}.
Note that $\d\theta$ is the counting measure on $\{\theta = \theta_0 \times 2^j : j\in\sdZ\}$. 
Consider setting $B = \{\theta_0\}$. \cref{assum:paramsupport} holds if
for all $x\in\scX$ and $m$-a.e. $z\in\scZ$,
\[
\int \1[\mu(x,z,a,b) = \theta_0 = \mu(f_{\theta_0}(x,z),a,b)]\1_\Delta(\d(a,b)) > 0.
\]
That is, if there is a nonzero probability of choosing the default parameter 
$\theta_0$ at any point $(x,z)$. Note that 
\[
  &\mu(x,z,a,b) = \theta_0 = \mu(f_{\theta_0}(x,z),a,b) \\
  &\iff \log(a) < \ell(x,z,\theta_0) < \log(b) \quad \text{or} \quad \log(a) < -\ell(x,z,\theta_0) < \log(b).
\]
By assumption, for all $x \in \scX$ and $m$-a.e. $z \in \scZ$, we have 
$\ell(x,z,\theta_0) \notin \cbra{-\infty, 0, \infty}$. 
If $\ell(x,z,\theta_0) > 0$, then when $a < \exp(-\ell(x,z,\theta_0)) < b$ we have 
the condition hold. This has positive measure under $\1_\Delta(\d (a,b))$. 
If $\ell(x,z,\theta_0) < 0$, then when $a < \exp(\ell(x,z,\theta_0)) < b$, the above condition holds.
This set also has positive measure.
\eprfof

\blem 
\label{lem:RWMH_irreducible}
For the AutoStep RWMH kernel with any fixed $\theta > 0$, $x \in \scX$, and $A \subset \scX$ with $\pi(A) > 0$,
we have $P_\theta(x, A) > 0$, provided that $\pi(x) > 0$ for all $x \in \scX$.
\elem 

\bprfof{\cref{lem:RWMH_irreducible}}
Fix $\theta > 0$, $A \subset \scX$ with $\pi(A) > 0$ and $x \in \scX$. 
Because $\pi \ll \lambda$, we have $\lambda(A) > 0$.
By translation properties of the Lebesgue measure, for any $x \in \reals^d$, 
$\lambda(A) = \lambda(A - x) > 0$, where $A-x = \cbra{\tilde x - x : \tilde x \in A}$.
Here, $(x',z') = f_\theta(x,z) = (x+z, -z)$ and so $\abs{\nabla f_\theta(x,z)} = 1$. 
Also, since $z \sim m$ where $m = \Norm(0, I)$, we have 
\[
  \ell(x,z,\theta) = \log\lt( \frac{\pi(x')}{\pi(x)}\rt).
\]
Then,
\[
  P_\theta(x, A) 
  \geq \int \1[z \in A-x] \min\lt\{1, \frac{\pi(x+z)}{\pi(x)}\rt\} m(z) \, \lambda(\d z) 
  > 0,
\]
because for all $x,z$ we have $\lambda(A-x) > 0$, $m(z) > 0$, and $\min\{1, \pi(x+z)/\pi(x)\} > 0$.
\eprfof

\blem
\label{lem:MALA_irreducible} 
For the AutoStep MALA kernel with any fixed $\theta > 0$, $x \in \scX$, 
differentiable $\pi$, positive definite $M$, and $A \subset \scX$ with $\pi(A) > 0$,
we have $P_\theta(x, A) > 0$, provided that $\pi(x) > 0$ for all $x \in \scX$.
\elem

\bprfof{\cref{lem:MALA_irreducible}} 
Fix $\theta > 0$, $A \subset \scX$ with $\pi(A) > 0$ and $x \in \scX$. 
Because $\pi \ll \lambda$, we have $\lambda(A) > 0$.
As in \cite{BironLattes24}, we combine the updates on $(x,z)$ into one step, so that 
$f_\theta(x,z) = (x'(\theta), z'(\theta))$, where
\[
  x'(\theta) &= x + \theta M^{-1} z + \frac{\theta^2}{2} M^{-1} \nabla \log \gamma(x), \qquad
  z'(\theta) = -\left(z + \frac{\theta}{2} \nabla \log \gamma(x) + 
    \frac{\theta}{2} \nabla \log \gamma(x'(\theta))\right).
\]
Now, $x'(\theta) \in A$ if 
\[
  z \in A_x := \lt\{\frac{M (\tilde x - x)}{\theta} - \frac{\theta}{2} \nabla \log \gamma(x) : \tilde x \in A \rt\}.
\]
By translation and scaling properties of the Lebesgue measure, for any $x \in \reals^d$, 
$\lambda(A_x) > 0$. 
It is a standard result that the leapfrog integrator satisfies 
$\abs{\nabla f_\theta(x,z)} = 1$. We have 
\[
  \ell(x,z,\theta) = \log\lt( \frac{\pi(x') m(z')}{\pi(x) m(z)} \rt).
\]
Then,
\[
  P_\theta(x, A) 
  \geq \int \1[z \in A_x] \min\lt\{1, \frac{\pi(x') m(z')}{\pi(x) m(z)} \rt\} m(z) \, \lambda(\d z) 
  > 0.
\]
because for all $x,z$ we have $\lambda(A_x) > 0$, $m(z) > 0$, and the acceptance ratio is positive. 
\eprfof

\bprfof{\cref{prop:stepsizeterminates}}
We generalize the step size termination proof of Theorem 3.3 by \citet{BironLattes24}.
Consider first the case where $v=-1$.
Since for $\pi\times m$-\aev $x,z$,  $\lim_{\theta \to 0^+} |\ell(x,z,\theta)| = 0$, and $|\log a| > 0$ almost surely,
there exists a $\theta' > 0$ such that $\forall 0 < \theta < \theta'$, $|\ell(x,z,\theta)| < |\log a|$.
Therefore there exists a $j < 0$ such that $2^j \theta_0 < \theta'$ and the while loop terminates.
Next consider the case where $v=1$. Since
for $\pi\times m$-\aev $x,z$,  $\lim_{\theta \to \infty} |\ell(x,z,\theta)| = \infty$, and $|\log b| < \infty$ almost surely,
there exists a $\theta' > 0$ such that $\theta > \theta'$, $|\ell(x,z,\theta)| > |\log b|$.
Therefore there exists a $j > 0$ such that $2^j \theta_0 > \theta'$ and the while loop terminates.

\eprfof

\bprfof{\cref{prop:expectedstepsizetermination}}
\[
\E\tau(s, \theta_0) &= \sum_{t=0}^\infty \P(\tau(s,\theta_0) > t)\\
&= \sum_{t=0}^\infty \P\lt( \max_{0\leq j \leq t} |\ell(x,z,2^j\theta_0)| < |\log b| \cup \min_{-t\leq j \leq 0} |\ell(x,z,2^{j}\theta_0)| > |\log a|\rt)\\
&\leq \sum_{t=0}^\infty \P\lt( \max_{0\leq j \leq t} |\ell(x,z,2^j\theta_0)| < |\log b|\rt) + \P\lt(\min_{-t\leq j \leq 0} |\ell(x,z,2^{j}\theta_0)| > |\log a|\rt)\\
&\leq \sum_{t=0}^\infty \P\lt( |\ell(x,z,2^t\theta_0)| < |\log b|\rt) + \P\lt(|\ell(x,z,2^{-t}\theta_0)| > |\log a|\rt)\\
&= \sum_{t=0}^\infty \P\lt( |\ell(x,z,2^t\theta_0)| < -\log b\rt) + \P\lt(|\ell(x,z,2^{-t}\theta_0)| > -\log a\rt)\\
&= \sum_{t=0}^\infty \P\lt( e^{-|\ell(x,z,2^t\theta_0)|} > b\rt) + \P\lt(e^{-|\ell(x,z,2^{-t}\theta_0)|} < a\rt).
\]
Since $a, b$ are uniform on $0 \leq a < b \leq 1$,
\[
\P(a > x) = (1-x)^2, \quad\text{and}\quad \P(b < x) = x^2,
\]
so
\[
\E\tau(s, \theta_0) &\leq \sum_{t=0}^\infty \E e^{-2|\ell(x,z,2^t\theta_0)|} + \E\lt[ \lt(1-e^{-|\ell(x,z,2^{-t}\theta_0)|}\rt)^2\rt], \quad (x,z)\dist \pi\times m.
\]
Fubini's theorem completes the proof.
\eprfof

\bprfof{\cref{cor:expectedtaurwmh}}
The involution for RWMH with $M=I$ is given by
\[
f_\theta(x,z) = (x+\theta z, -z), \quad m=\scN(0, I).
\]
Therefore,
\[
\ell(x,z,\theta) = \log\lt(\frac{\pi(x+\theta z), m(-z)}{\pi(x) m(z)}\rt) = \log \pi(x+\theta z) - \log \pi(x).
\]
Since $\log \pi$ is $L$-Lipschitz smooth,
\[
|\ell(x,z,\theta)| &\leq \theta |\grad \log \pi(x)^Tz| + \frac{1}{2}L\theta^2 \|z\|^2.
\]
Therefore,
\[
\sum_{t=0}^\infty \lt(1-e^{-|\ell(x,z,2^{-t}\theta_0)|}\rt)^2 
&\leq \sum_{t=0}^\infty \lt(1-e^{-2^{-t}\theta_0 |\grad \log \pi(x)^Tz| - \frac{1}{2}L\theta_0^2 4^{-t} \|z\|^2}\rt)^2\\
&= \sum_{t=0}^\infty 1 - 2e^{-2^{-t}\theta_0 |\grad \log \pi(x)^Tz| - \frac{1}{2}L 4^{-t}\theta_0^2 \|z\|^2} + e^{-2\cdot 2^{-t}\theta_0 |\grad \log \pi(x)^Tz| - L4^{-t}\theta_0^2 \|z\|^2}.
\]

Furthermore, note that $(1-e^{-x})^2 \leq \min(x^2,1)$ for $x\geq 0$. Therefore,
\[
\sum_{t=0}^\infty \lt(1-e^{-|\ell(x,z,2^{-t}\theta_0)|}\rt)^2 
&\leq \sum_{t=0}^\infty\min\lt(\ell(x,z,2^{-t}\theta_0)^2, 1\rt)\\
&\leq \sum_{t=0}^\infty\min\lt(\lt(2^{-t}\theta_0 |\grad \log \pi(x)^Tz| + \frac{1}{2}L4^{-t}\theta_0^2 \|z\|^2\rt)^2, 1\rt)\\
&\leq \sum_{t=0}^\infty\min\lt(4^{-t}\lt(\theta_0 |\grad \log \pi(x)^Tz| + \frac{1}{2}L\theta_0^2 \|z\|^2\rt)^2, 1\rt).
\]
Let $t_0 = \lceil \max\{0, (3/2)\log\lt(\theta_0 |\grad \log \pi(x)^Tz| + \frac{1}{2}L\theta_0^2 \|z\|^2\rt)\} \rceil$.
Then,
\[
\sum_{t=0}^\infty \lt(1-e^{-|\ell(x,z,2^{-t}\theta_0)|}\rt)^2 
&\leq \sum_{t=0}^{t_0-1} 1 + \sum_{t=t_0}^\infty4^{-t}\lt(\theta_0 |\grad \log \pi(x)^Tz| + \frac{1}{2}L\theta_0^2 \|z\|^2\rt)^2\\
&= t_0 + \sum_{t=0}^\infty4^{-t} 4^{-t_0}\lt(\theta_0 |\grad \log \pi(x)^Tz| + \frac{1}{2}L\theta_0^2 \|z\|^2\rt)^2\\
&\leq t_0 + \sum_{t=0}^\infty 4^{-t} \leq t_0 + 4/3.
\]

Next, since $\log \pi$ is $U$-strongly log-concave,
\[
\ell(x,z,\theta) &\leq \theta \grad\log\pi(x)^Tz - \frac{1}{2}U\theta^2\|z\|^2.
\]
If $\grad\log\pi(x)^Tz \leq 0$, then
\[
|\ell(x,z,\theta)| &\geq \theta |\grad\log\pi(x)^Tz| + \frac{1}{2}U\theta^2\|z\|^2 = \frac{1}{2}U\theta^2\|z\|^2 - \theta \grad\log\pi(x)^Tz. 
\]
On the other hand, if $\grad\log\pi(x)^Tz > 0$, then
\[
|\ell(x,z,\theta)| &\geq \lt\{\begin{array}{ll}
0 & \theta \leq \frac{2 \grad\log\pi(x)^Tz}{U\|z\|^2}\\
\frac{1}{2}U\theta^2\|z\|^2 - \theta \grad\log\pi(x)^Tz  & \theta > \frac{2 \grad\log\pi(x)^Tz}{U\|z\|^2}
\end{array}. \rt.
\]
Taken together, for $\theta \geq 0$, 
\[
|\ell(x,z,\theta)| &\geq \1\lt[\theta > \frac{2 \grad\log\pi(x)^Tz}{U\|z\|^2}\rt]\lt(\frac{1}{2}U\theta^2\|z\|^2 - \theta \grad\log\pi(x)^Tz\rt)\\
&\geq \1\lt[\theta > \frac{2 \grad\log\pi(x)^Tz+1}{U\|z\|^2}\rt]\lt(\frac{1}{2}U\theta^2\|z\|^2 - \theta \grad\log\pi(x)^Tz\rt).
\]
So,
\[
\sum_{t=0}^\infty e^{-2|\ell(x,z,2^t\theta_0)|}
&\leq\sum_{t=0}^\infty e^{-2\cdot 2^t\theta_0 \1\lt[2^t\theta_0 > \frac{2 \grad\log\pi(x)^Tz+1}{U\|z\|^2}\rt]\lt(\frac{1}{2}U 2^t \theta_0\|z\|^2 - \grad\log\pi(x)^Tz\rt)}\\
&\leq\sum_{t=0}^\infty e^{-2\cdot 2^t\theta_0 \1\lt[2^{t/2}\theta_0 > 1, \, 2^t\theta_0 > \frac{2 \grad\log\pi(x)^Tz+1}{U\|z\|^2}\rt]\lt(\frac{1}{2}U 2^t \theta_0\|z\|^2 - \grad\log\pi(x)^Tz\rt)}.
\]
Let $t_0' = \lceil \max\{0, -3\log\theta_0,  (3/2)\log \frac{2 \grad\log\pi(x)^Tz+1}{U\theta_0\|z\|^2}\}\rceil$, 
where $\log$ of a negative number is taken to be $-\infty$. Then,
\[
\sum_{t=0}^\infty e^{-2|\ell(x,z,2^t\theta_0)|}
&\leq \sum_{t=0}^{t'_0-1}1 + \sum_{t=t'_0}^\infty e^{-2\cdot 2^t \theta_0 \1\lt[\dots\rt]\lt(\frac{1}{2}U2^t \theta_0\|z\|^2 - \grad\log\pi(x)^Tz\rt)}\\
&\leq \sum_{t=0}^{t'_0-1}1 + \sum_{t=t'_0}^\infty e^{-2\cdot 2^{t/2} }\\
&\leq t'_0 + \sum_{t=0}^\infty e^{-2^{(t+1)/2}}\\
&\leq t'_0 + \int_0^\infty e^{-2^{t/2}} \, \d t\\
&= t'_0 + \frac{2}{\log 2}\int_1^\infty t^{-1}e^{-t} \, \d t\\
&\leq t'_0 + \frac{3}{4}.
\]
Combining the above two bounds, as well as the bounds $\lceil x \rceil \leq x+1$ for $x > 0$, $u^Tv \leq \|u\|\|v\|$ for vectors $u,v$,
and $\|\grad \log \pi(x)\| \leq L\|x\|$ for an $L$-Lipschitz smooth function (without loss of generality, assume $\pi(x)$ reaches its maximum and
has $\grad \log \pi(x) = 0$ at $x=0$): 
\[
\E\tau(s,\theta_0) &\leq \E t_0 + \E t'_0 + \frac{4}{3} + \frac{3}{4}\\
&\leq 5 + 3\log\lt(1+\theta_0^{-1}\rt) +\frac{3}{2}\E \log\lt(1+\theta_0 L\|x\|\|z\| + \frac{1}{2}L\theta_0^2 \|z\|^2\rt)
+ \frac{3}{2}\E \log \lt(1+\frac{2 L\|x\|\|z\|+1}{U\theta_0\|z\|^2}\rt).
\]
Jensen's inequality yields
\[
\E\tau(s,\theta_0)
&\leq 5 + 3\log\lt(1+\theta_0^{-1}\rt)  +\frac{3}{2} \log\lt(1+\theta_0 L\E\|x\|\E\|z\| + \frac{1}{2}L\theta_0^2 \E\|z\|^2\rt)
+ \frac{3}{2}\E \log \lt(1+\frac{2 L\|z\|\E\|x\|+1}{U\theta_0\|z\|^2}\rt).
\]
Since $z \dist\Norm(0,I)$ in $\reals^d$,
\[
\E\|z\| &= \sqrt{2}\frac{\Gamma\lt(\frac{d+1}{2}\rt)}{\Gamma\lt(\frac{d}{2}\rt)} \leq \lt(d+1\rt)^{1/2}, \quad\text{and}\quad \E\|z\|^2  = d,
\]
where the bound on $\E\|z\|$ follows from Gautschi's inequality.
Since $x$ has a strongly log-concave and Lipschitz smooth density,
\[
\E\|x\| &= \frac{\int e^{\log \pi(x)} \|x\| \, \d x}{\int e^{\log \pi(x)} \, \d x} 
\leq  \frac{(2\pi U^{-1})^{d/2}\int \frac{1}{(2\pi U^{-1})^{d/2}}e^{-\frac{1}{2}U\|x\|^2} \|x\| \, \d x}{(2\pi L^{-1})^{d/2}\int \frac{1}{(2\pi L^{-1})^{d/2}} e^{-\frac{1}{2}L\|x\|^2}\, \d x} 
= \frac{U^{-(d+1)/2}\E\|z\|}{L^{-d/2}} \leq \frac{L^{d/2}(d+1)^{1/2}}{U^{(d+1)/2}}. 
\]
Finally, for any $a, b > 0$,
\[
\E \log \lt(1 + \frac{a \|z\| + 1}{b \|z\|^2}\rt) 
&= \E \1[\|z\| < 1]\log \lt(1 + \frac{a \|z\| + 1}{b \|z\|^2}\rt) + \E \1[\|z\| \geq 1]\log \lt(1 + \frac{a \|z\| + 1}{b \|z\|^2}\rt)\\
&\leq \E \1[\|z\| < 1]\log \lt(1 + \frac{a + 1}{b \|z\|^2}\rt) + \E \1[\|z\| \geq 1]\log \lt(1 + \frac{a \|z\| + 1}{b}\rt)\\
&\leq \E\1[\|z\| < 1]\log \lt(1 + \frac{a + 1}{b \|z\|^2}\rt) + \log \lt(1 + \frac{a \E \|z\| + 1}{b}\rt)\\
&= \int_0^\infty \P\lt(\1[\|z\| < 1]\log \lt(1 + \frac{a + 1}{b \|z\|^2}\rt) > t\rt)\d t + \log \lt(1 + \frac{a \E \|z\| + 1}{b}\rt)\\
&= \int_0^\infty \P\lt(\|z\|^2 < \min\lt\{1, \frac{a+1}{b(e^t-1)}\rt\}\rt)\d t + \log \lt(1 + \frac{a \E \|z\| + 1}{b}\rt)\\
&= \int_0^\infty \frac{1}{2^{d/2}\Gamma(d/2)}\int_0^{\min\lt\{1, \frac{a+1}{b(e^t-1)}\rt\}} s^{d/2-1} e^{-s/2}\d s \, \d t + \log \lt(1 + \frac{a \E \|z\| + 1}{b}\rt)\\
&\leq \int_0^\infty \frac{1}{2^{d/2}\Gamma(d/2)}\int_0^{\min\lt\{1, \frac{a+1}{b(e^t-1)}\rt\}} s^{d/2-1}\d s \, \d t + \log \lt(1 + \frac{a (d+1)^{1/2} + 1}{b}\rt)\\
&= \int_0^\infty \frac{2}{d2^{d/2}\Gamma(d/2)}\lt(\min\lt\{1, \frac{a+1}{b(e^t-1)}\rt\}\rt)^{d/2} \d t + \log \lt(1 + \frac{a (d+1)^{1/2}+ 1}{b}\rt)\\
&= \frac{2}{d2^{d/2}\Gamma(d/2)}\lt(\int_0^{\log\lt(1+\frac{a+1}{b}\rt)} \d t + \int_{\log\lt(1+\frac{a+1}{b}\rt)}^\infty \lt(\frac{a+1}{b(e^t-1)}\rt)^{d/2} \d t\rt) \\
& \quad + \log \lt(1 + \frac{a (d+1)^{1/2}+ 1}{b}\rt)\\
&= \frac{2}{d2^{d/2}\Gamma(d/2)}\lt(\log\lt(1+\frac{a+1}{b}\rt) + \lt(\frac{a+1}{b}\rt)^{d/2}\int_{\log\lt(1+\frac{a+1}{b}\rt)}^\infty (e^t-1)^{-d/2} \d t\rt) \\
& \quad + \log \lt(1 + \frac{a (d+1)^{1/2}+ 1}{b}\rt)\\
\displaybreak
&= \frac{2}{d2^{d/2}\Gamma(d/2)}\lt(\log\lt(1+\frac{a+1}{b}\rt) + \lt(\frac{a+1}{b}\rt)^{d/2}\int_{\log\lt(\frac{a+1}{b}\rt)}^\infty e^{-sd/2}\frac{e^s}{1+e^s} \d s\rt) \\
& \quad + \log \lt(1 + \frac{a (d+1)^{1/2}+ 1}{b}\rt)\\
&\leq \frac{2}{d2^{d/2}\Gamma(d/2)}\lt(\log\lt(1+\frac{a+1}{b}\rt) + \lt(\frac{a+1}{b}\rt)^{d/2}\int_{\log\lt(\frac{a+1}{b}\rt)}^\infty e^{-sd/2} \d s\rt) \\
& \quad + \log \lt(1 + \frac{a (d+1)^{1/2}+ 1}{b}\rt)\\
&= \frac{2}{d2^{d/2}\Gamma(d/2)}\lt(\log\lt(1+\frac{a+1}{b}\rt) + \lt(\frac{a+1}{b}\rt)^{d/2}\frac{2}{d} \lt(\frac{a+1}{b}\rt)^{-d/2}\rt) \\
& \quad + \log \lt(1 + \frac{a (d+1)^{1/2}+ 1}{b}\rt)\\
&= \frac{2}{d2^{d/2}\Gamma(d/2)}\lt(\log\lt(1+\frac{a+1}{b}\rt) + \frac{2}{d}\rt)  + \log \lt(1 + \frac{a (d+1)^{1/2}+ 1}{b}\rt).
\]
Therefore, the final bound is
\[
\E\tau(s, \theta_0) &\leq
5 + 3\log\lt(1+\theta_0^{-1}\rt)  +\frac{3}{2} \log\lt(1+\theta_0 L\frac{L^{d/2}(d+1)^{1/2}}{U^{(d+1)/2}}(d+1)^{1/2} + \frac{1}{2}L\theta_0^2 d\rt) \\
& \quad + \frac{3}{2}\E \log \lt(1+\frac{2 L\|z\|\frac{L^{d/2}(d+1)^{1/2}}{U^{(d+1)/2}}+1}{U\theta_0\|z\|^2}\rt)\\
&\leq
5 + 3\log\lt(1+\theta_0^{-1}\rt)  +\frac{3}{2} \log\lt(1+\theta_0 L\frac{L^{d/2}(d+1)}{U^{(d+1)/2}} + \frac{1}{2}L\theta_0^2 d\rt)\\
& \quad + \frac{3}{2}\frac{2}{d2^{d/2}\Gamma(d/2)}\lt(\log\lt(1+\frac{2 L\frac{L^{d/2}(d+1)^{1/2}}{U^{(d+1)/2}}+1}{U\theta_0}\rt) + \frac{2}{d}\rt)  + \frac{3}{2}\log \lt(1 + \frac{2 L\frac{L^{d/2}(d+1)}{U^{(d+1)/2}} + 1}{U\theta_0}\rt)\\
&\leq
5 + 3\log\lt(1+\theta_0^{-1}\rt)+ \frac{6}{2^{d/2}}   +\frac{3}{2} \log\lt(1+\theta_0 L\frac{L^{d/2}(d+1)}{U^{(d+1)/2}} + \frac{1}{2}L\theta_0^2 d\rt) \\
& \quad + \frac{3}{2}\lt(\frac{2}{2^{d/2}}+1\rt)\log\lt(1+\frac{2 L\frac{L^{d/2}(d+1)}{U^{(d+1)/2}}+1}{U\theta_0}\rt).
\]
The result follows by inspection of the above bound.
\eprfof


\bprfof{\cref{prop:energyjumpdist}}
The expected jump distance is
\[
  \E D &= \int |\ell(x,z,\theta)| \min\lt\{1, e^{\ell(x,z,\theta)}\frac{\eta(\theta|f_\theta(x,z),a,b)}{\eta(\theta|x,z,a,b)}\rt\} \pi(\d x)m(\d z)(2\1_\Delta(\d(a,b)))\eta(\d\theta \mid x,z,a,b).
\]
Let $A_{\leq} = \{x,z,a,b,\theta : \exp(\ell(x,z,\theta)) \leq \eta(\theta|x,z,a,b) / \eta(\theta|f_\theta(x,z),a,b)\}$, and
define $A_<, A_{\geq}, A_>$ similarly by replacing the inequality in the definition.
We begin by splitting the integral into regions $A_{\leq}, A_{>}$:
\[
  \E D &= \int_{A_{>}} |\ell(x,z,\theta)| \pi(\d x)m(\d z)(2\1_\Delta(\d(a,b)))\eta(\d\theta \mid x,z,a,b)\\
  +&  \int_{A_\leq} |\ell(x,z,\theta)| e^{\ell(x,z,\theta)}\eta(\theta|f_\theta(x,z),a,b)\pi(\d x)m(\d z)(2\1_\Delta(\d(a,b)))\d\theta.
\]
Next we perform the transformation of variables $x', z' = f_\theta(x,z)$ on the $A_{>}$ integral.
Recall that $\ell(x,z,\theta)$ satisfies
\[
  \ell(f_\theta(x,z), \theta) = -\ell(x,z,\theta),
\]
and note that this transformation yields the integration region $A_{<}$, such that
\[
  \E D &=  \int_{A_<} |\ell(x,z,\theta)|e^{\ell(x,z,\theta)} \eta(\theta|f_\theta(x,z),a,b)\pi(\d x)m(\d z)(2\1_\Delta(\d(a,b)))\d\theta\\
  +&  \int_{A_\leq} |\ell(x,z,\theta)| e^{\ell(x,z,\theta)}\eta(\theta|f_\theta(x,z),a,b)\pi(\d x)m(\d z)(2\1_\Delta(\d(a,b)))\d\theta.
\]
The upper bound follows by combining the two terms and bounding the $\eta$ ratio:
\[
  \E D &\leq 2\int_{A_\leq} |\ell(x,z,\theta)|e^{\ell(x,z,\theta)} \frac{\eta(\theta|f_\theta(x,z),a,b)}{\eta(\theta|x,z,a,b)}\eta(\d\theta|x,z,a,b)\pi(\d x)m(\d z)(2\1_\Delta(\d(a,b)))\\
   &\leq 2\sbeta \sup_{y\leq \log\sbeta} |y| e^{y} \sbpi(A_\leq)\\
  &\leq 2\sbeta \sup_{y\leq \log\sbeta} |y| e^{y}\\
  &= 2\sbeta\max\lt\{e^{-1}, \sbeta\log\sbeta \rt\}.
\]
\eprfof

\newpage
\section{Additional Experimental Details and Results} 
\label{app:experiments}

\subsection{Synthetic and real data models}
We present the synthetic and real data models used in \cref{sec:experiments}. In the
following, $\Norm(\mu, \sigma^2)$ represents a normal distribution
with mean $\mu$ and variance $\sigma^2$.  We use 
$\Norm(\mu, \sigma^2; b_L, b_U)$ to denote a truncated normal distribution with 
bounds $(b_L, b_U)$.

\begin{itemize}
\item The Neal's funnel with $d$ dimensions ($d \ge 2$) and scale parameter $\tau > 0$, denoted by funnel($d, \tau$), is given by
\[
X_1 \sim \Norm(0, 9), \quad X_2, \dots, X_d \mid X_1 = x_1 \distiid \Norm(0, (\exp(x_1 / \tau))^2).
\]
For the 2-dimensional funnel we used $\tau = 0.6$, and for the 100-dimensional funnel we used $\tau = 6$.

\item The mRNA model with $N$ observations, the time $t \in \mathbb{R}^N$, and the observed outcomes $y \in \mathbb{R}^N$ is given by
\[
    \log_{10}(t_0) & \sim \text{Uniform}(-2, 1)\\
    \log_{10}(k_0) & \sim \text{Uniform}(-5, 5)\\
    \log_{10}(\beta) & \sim \text{Uniform}(-5, 5)\\
    \log_{10}(\delta) & \sim \text{Uniform}(-5, 5)\\
    \log_{10}(\sigma) & \sim \text{Uniform}(-2, 2)\\
    \mu_i & =
    \begin{cases} 
    0, & t_i - t_0 \leq 0 \\
    k_0 \cdot \dfrac{\exp(-\beta \cdot (t_i - t_0)) - \exp(-\delta \cdot (t_i - t_0))}{\delta - \beta}, & \text{if } \delta \neq \beta \\
    k_0 \cdot (t_i - t_0), & \text{if } \delta = \beta 
    \end{cases} \quad, i = 1, \dots, N \\
    y_i \mid \mu_i, \sigma & \sim \Norm(\mu_i, \sigma^2), \quad i = 1, \dots, N.
\]
\item The kilpisjarvi model with the predictors $x \in \mathbb{R}^N$, the observed responses $y \in \mathbb{R}^N$, and additional parameters $\mu_\alpha, \mu_\beta, \sigma_\alpha, \sigma_\beta$, is given by
\[
    \alpha & \sim \Norm(\mu_\alpha, \sigma_\alpha^2)\\
    \beta & \sim \Norm(\mu_\beta, \sigma_\beta^2)\\
    \sigma & \sim \Norm(0, 1; 0, \infty)\\
    y_i \mid \alpha, \beta, \sigma, x_i & \sim \Norm(\alpha + \beta x_i, \sigma^2), \quad i = 1, \dots, N.
\]
In our experiments we used $\mu_\alpha = 9.313$, $\mu_\beta = 0$, $\sigma_\alpha = 100$, and $\sigma_\beta = 0.0333$.
\item The orbital model is a model of the multiple system Gliese 229 available in Octofitter.jl \citep{Octofitter}
at \url{https://github.com/sefffal/OrbitPosteriorDB/blob/main/models/astrom-GL229A.jl}.

\end{itemize}

\begin{table}[t]
    \centering
    \begin{tabular}{l|l|l}
    \hline
    \textbf{Model}                 & \textbf{Dimension} & $\alpha$      \\ \hline
    mRNA                           & 5                  & 5.767       \\ \hline
    orbital                        & 12                 & 5.389       \\ \hline
    kilpisjarvi                    & 3                  & 5.9561       \\ \hline
    funnel2                        & 2                  & 5.960         \\ \hline
    funnel100                      & 100                & 72.551      \\ \hline
    \end{tabular}
    \caption{The scaling constant $\alpha$ for each model, as mentioned in \cref{sec:experiments}.}
    \label{tab:alpha}
\end{table}

\subsection{Estimation of scaling factors}
\label{sec:scaling_factors}

We measure the cost of each method by $N_\ell+\alpha N_g$, where $N_\ell$ and $N_g$ 
are the number of evaluations of $\log\pi(\cdot)$ and $\grad\log\pi(\cdot)$, 
respectively. We use $\alpha$ as a target-dependent scaling 
factor to balance the cost of gradient and density evaluations. To determine $\alpha$, 
we ran three separate Markov chains using Pigeons to obtain draws, benchmarked the 
time required for gradient computations, and benchmarked the time for log density 
evaluations. The scaling factor $\alpha$ is then obtained as the ratio of the total 
time spent on gradient evaluations to the total time spent on log density evaluations.
The chains are long enough such that the estimate error (in absolute difference) 
is within 2\%. The $\alpha$ values used in all experiments are presented in 
\cref{tab:alpha}. 

There seems to be a positive correlation between $\alpha$ and the dimension of the 
problems. This correlation is likely due to the auto-differentiation system used to 
compute gradients. While auto-differentiation avoids the need for manual differentiation, 
it appears to incur a computational cost that increases with dimensionality.

\subsection{Algorithm Tuning Parameters}
We provide the detailed tuning parameters for the samplers compared in \cref{sec:experiment_compare_existing}. 

\begin{itemize}
    \item \textbf{Adaptive MALA}: Step size \( \theta_0 = 0.1 \).
    \item \textbf{Delayed Rejection HMC}: Step size \( \theta_0 = 0.1 \), number of subsequent proposals \( k = 2 \), and step size divisor \( a = 5.0 \).
    \item \textbf{NUTS}: Maximum tree depth \( L_{\max} = 5 \) and target acceptance ratio \( \alpha_{\text{target}} = 0.65 \).
\end{itemize}

\subsection{Additional results}

\begin{figure}[t!]
\centering \includegraphics[width=0.4\textwidth]{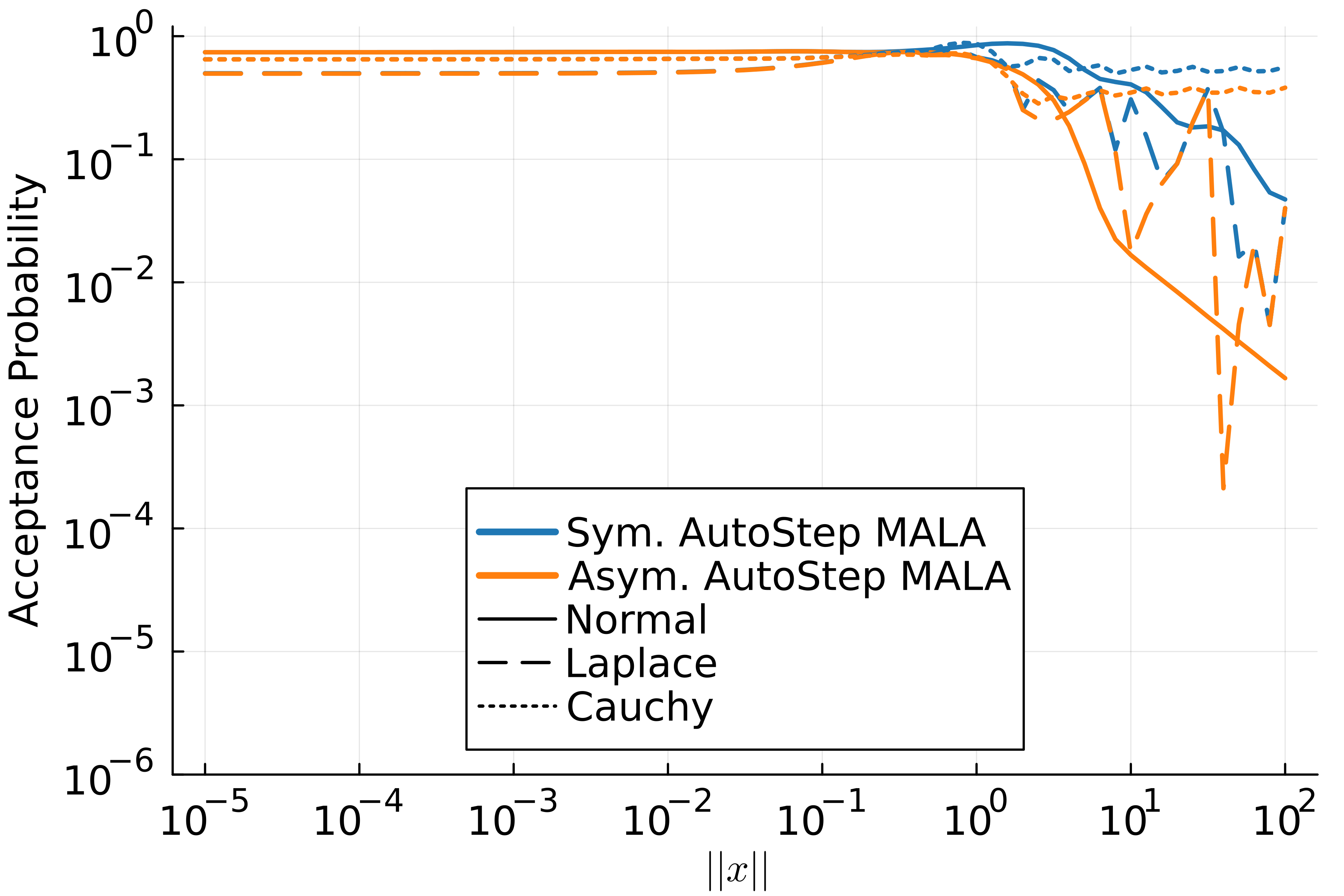}
	\caption{The same comparison as \ref{fig:sym_vs_asym}, except for the MALA involution
	instead of the RWMH involution. }\label{fig:sym_vs_asym_mala}
\end{figure}

\begin{figure}[t]
\begin{subfigure}{0.32\textwidth}
    \centering \includegraphics[width=\columnwidth]{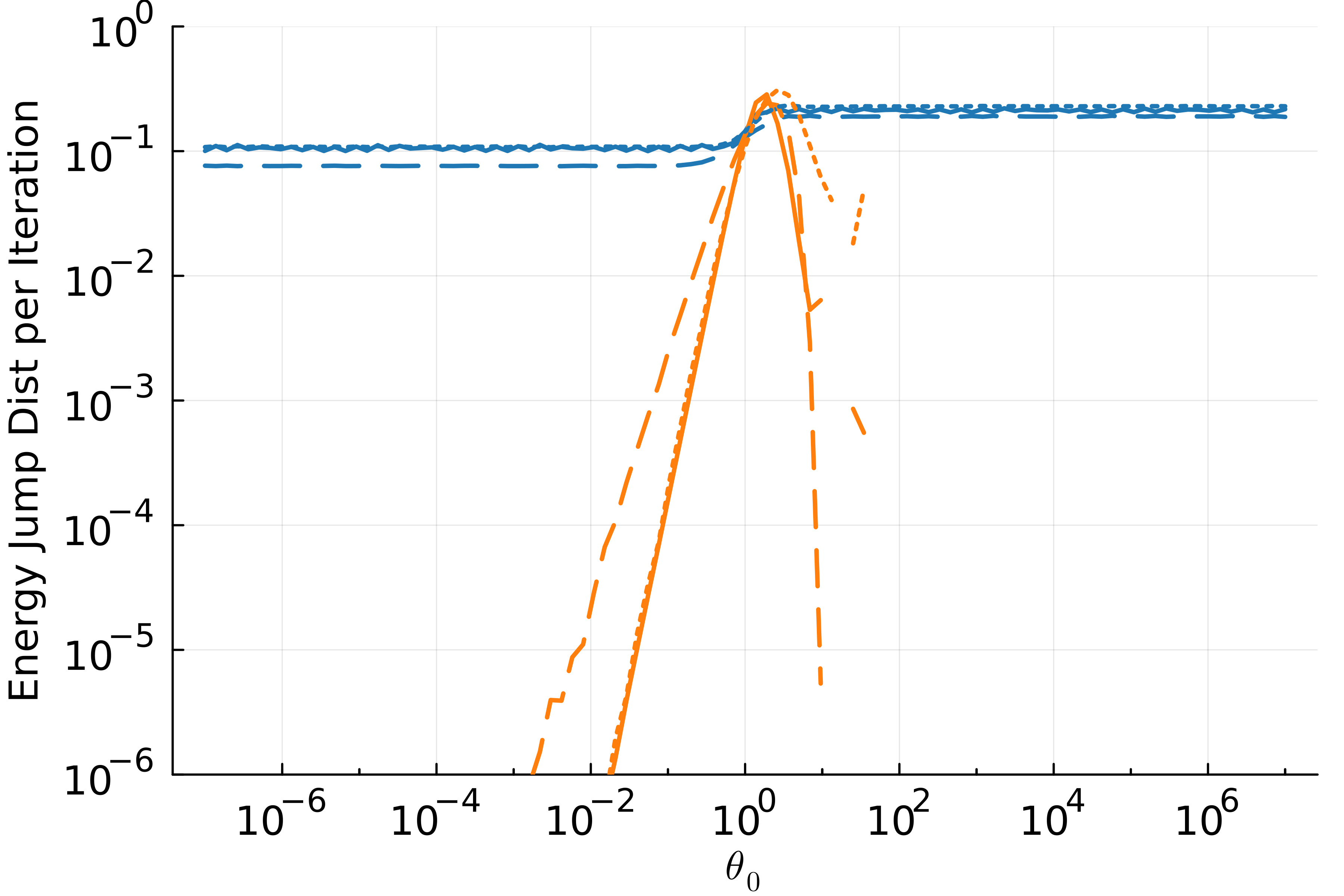}
   \centering\includegraphics[width=\columnwidth]{deliverables/ksess_and_ejump/legend_horizontal_MALA.png}
    \caption{Energy Jump Distance per Iteration}\label{fig:ejump_mala}
\end{subfigure}
\hfill
\begin{subfigure}{0.32\textwidth}
    \centering\includegraphics[width=\columnwidth]{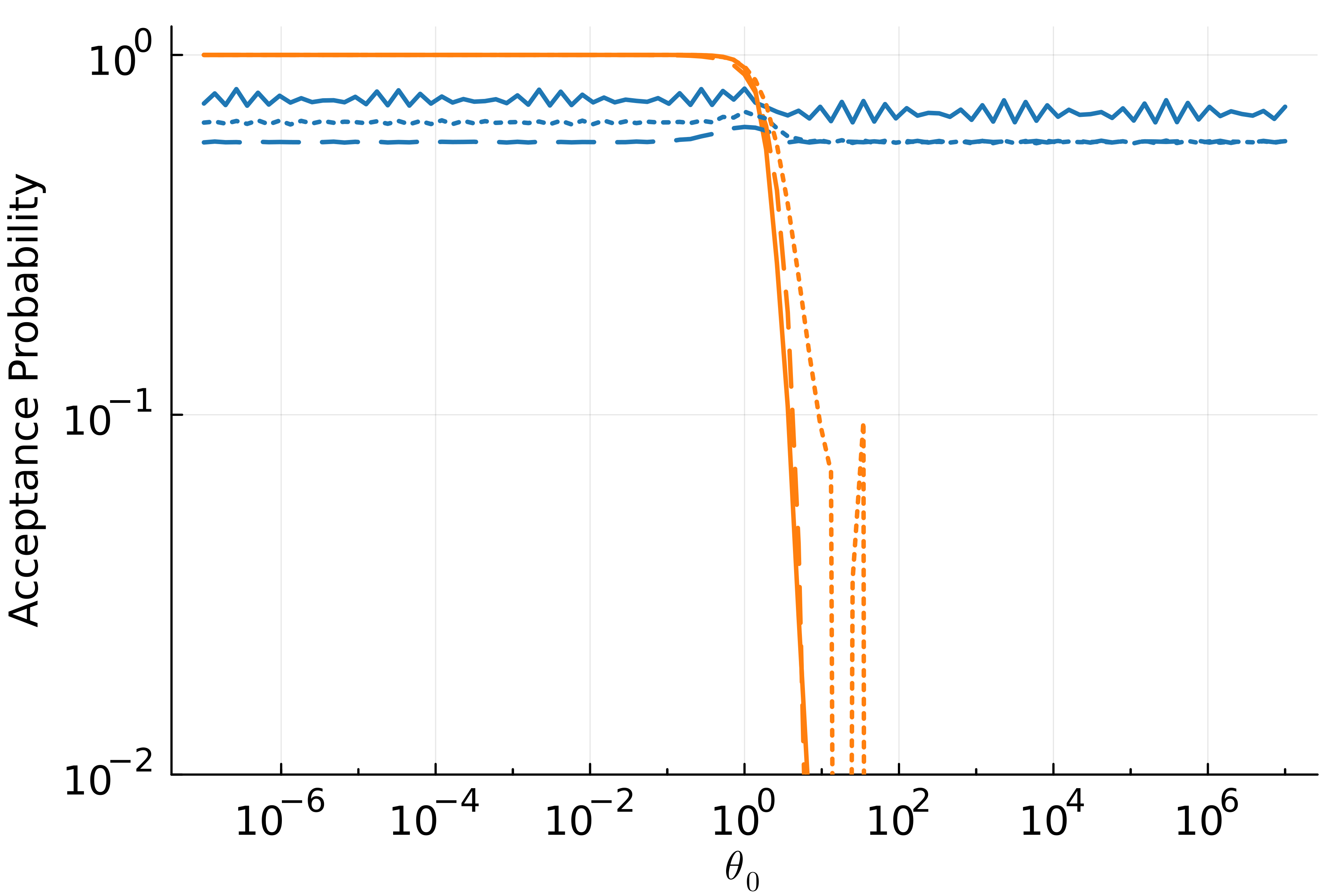}
   \centering\includegraphics[width=\columnwidth]{deliverables/ksess_and_ejump/legend_horizontal_MALA.png}
    \caption{Acceptance Probability}\label{fig:acc_mala}
\end{subfigure}
\hfill
\begin{subfigure}{0.32\textwidth}
    \centering\includegraphics[width=\columnwidth]{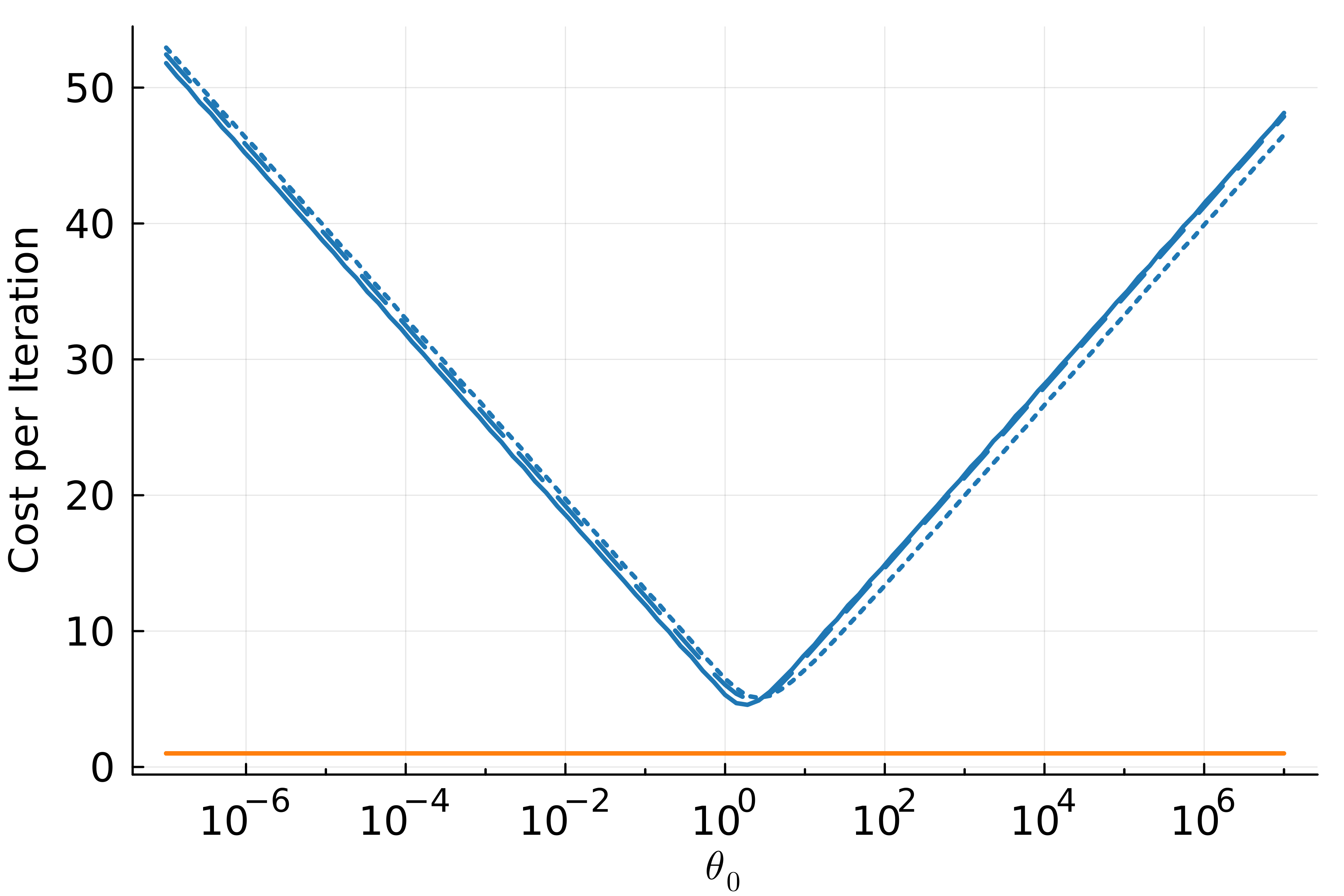}
   \centering\includegraphics[width=\columnwidth]{deliverables/ksess_and_ejump/legend_horizontal_MALA.png}
    \caption{Cost per Iteration}\label{fig:cost_mala}
\end{subfigure}
\caption{The same metrics as presented in \cref{fig:fixvsauto_other}, 
except for the MALA involution instead of the RWMH involution.}\label{fig:fixvsauto_other_mala}
\end{figure}

\begin{figure}[t]
\centering
\begin{subfigure}{0.32\textwidth}
    \centering \includegraphics[width=\columnwidth]{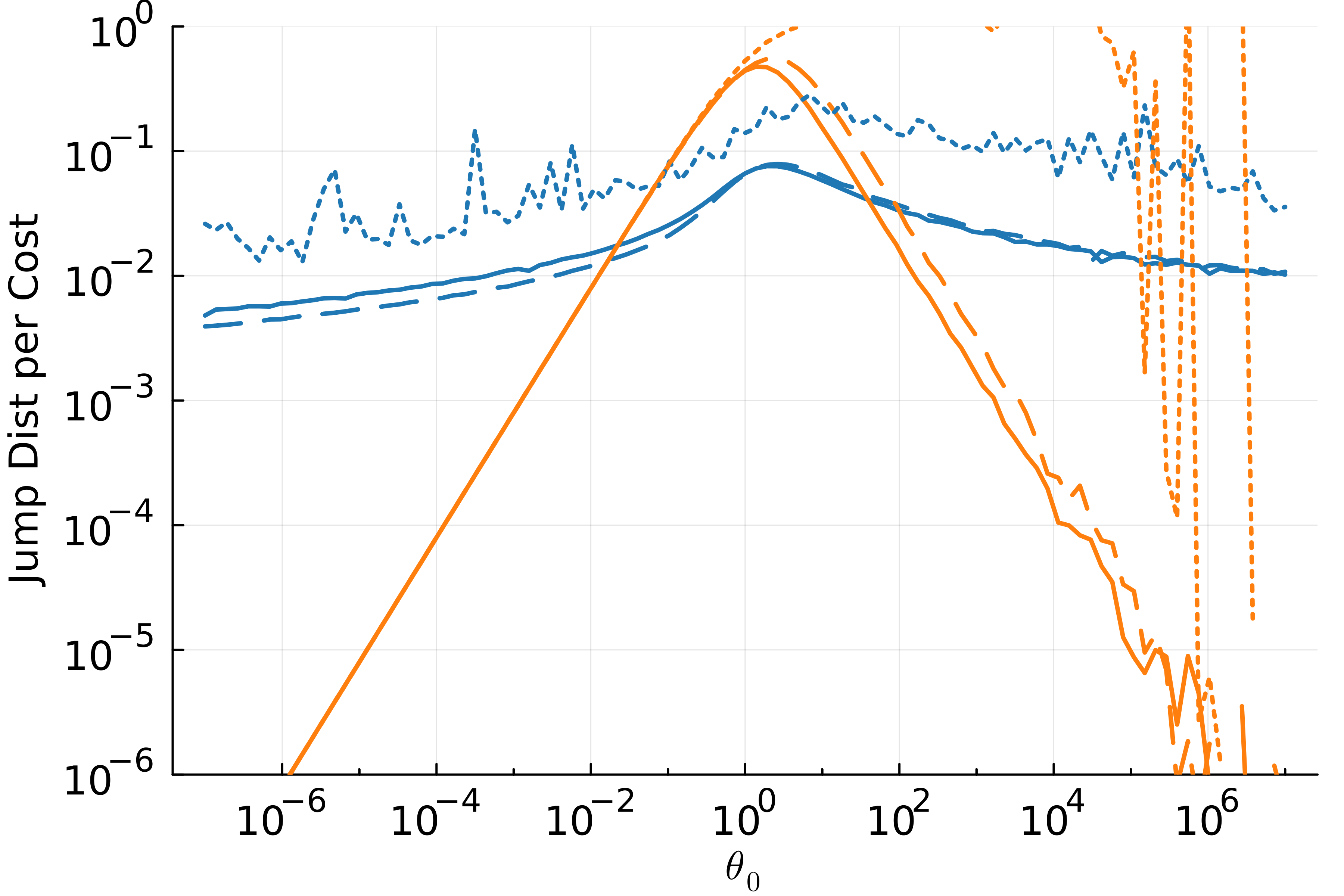}
    \centering\includegraphics[width=\columnwidth]{deliverables/ksess_and_ejump/legend_horizontal_RWMH.png}
    \caption{RWMH}\label{fig:jumprwmh}
\end{subfigure}
\begin{subfigure}{0.32\textwidth}
    \centering \includegraphics[width=\columnwidth]{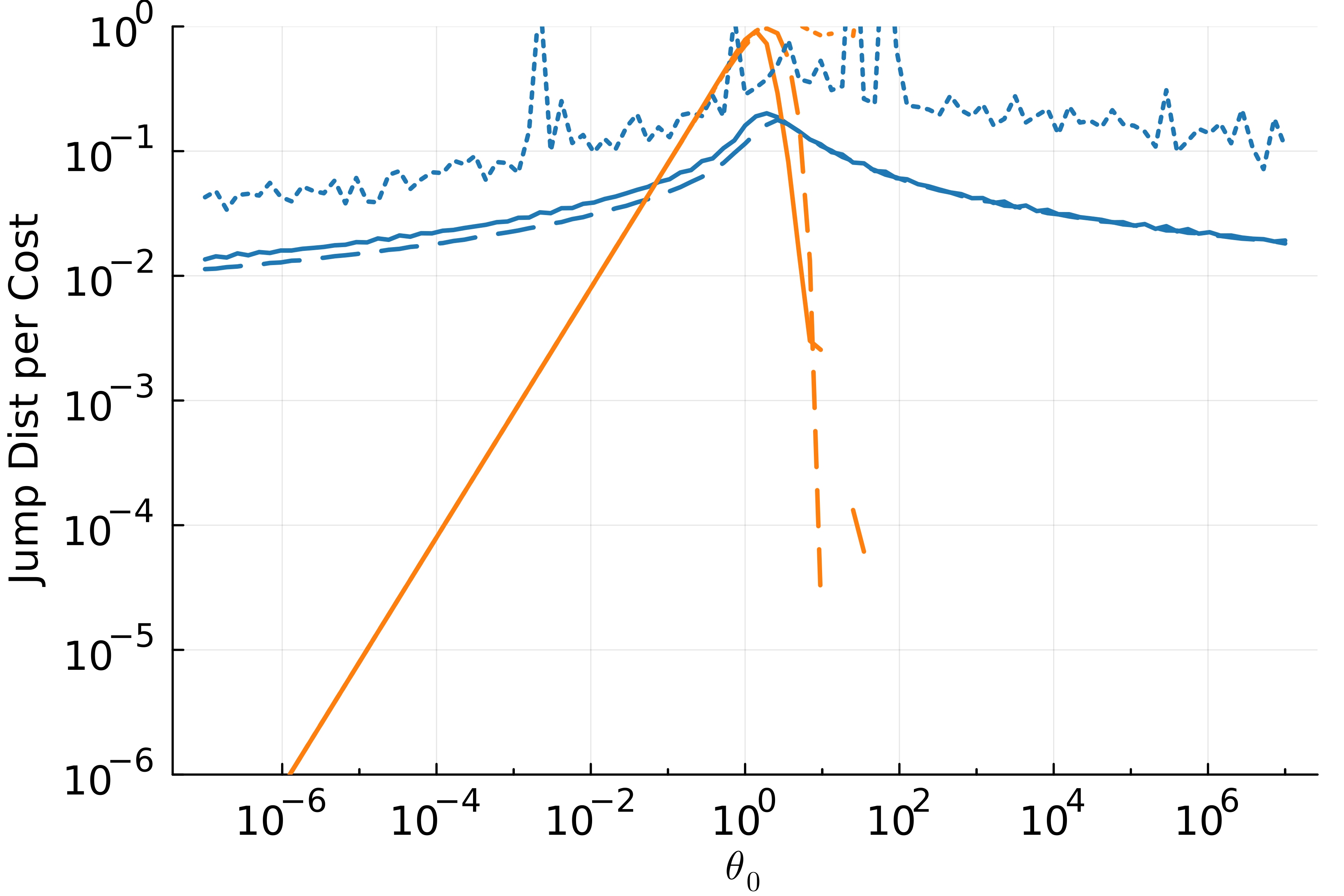}
    \centering\includegraphics[width=\columnwidth]{deliverables/ksess_and_ejump/legend_horizontal_MALA.png}
    \caption{MALA}\label{fig:jumpmala}
\end{subfigure}
\caption{The jump distance per iteration
for AutoStep and fixed step RWMH (\cref{fig:jumprwmh}) and MALA (\cref{fig:jumpmala}) versus initial step size $\theta_0$.}\label{fig:fixvsauto_jump}
\end{figure}

Additional results for the experiments in \cref{sec:expt_robustnesstheta0,sec:expt_symvsasym}
are presented in \cref{fig:sym_vs_asym_mala,fig:fixvsauto_other_mala,fig:fixvsauto_jump}.
In particular, \cref{fig:sym_vs_asym_mala} shows the same comparison of symmetric and asymmetric
step size selectors as in \cref{fig:sym_vs_asym} in the main text, except for the MALA involution.
The conclusion here is slightly different for MALA than for RWMH; 
both asymmetric and symmetric step size selection yields reasonable acceptance probabilities near the mode,
while both have decaying acceptance probabilities in the tails. However, despite the decay,
the symmetric criterion still yields a substantial increase over the asymmetric criterion, and should still
be preferred.
\cref{fig:fixvsauto_other_mala} shows the same additional metrics as \cref{fig:fixvsauto_other} when
comparing fixed step sizes vs AutoStep (energy jump distance per iteration,
acceptance probability, and cost per iteration) but for the MALA involution. The conclusions for MALA drawn here are precisely
the same as those for RWMH. Finally, \cref{fig:fixvsauto_jump} shows the expected jump distance for
fixed and AutoStep RWMH and MALA; these plots show the expected decay of at least $e^{-|\log\theta_0|}$ for fixed
step size methods, and $|\log\theta_0|^{-1}$ for AutoStep methods. It is worth pointing out that
the expected jump distance for fixed step size RWMH on very heavy tailed targets (like the Cauchy) can be quite high,
but indicates a few very large jumps rather than good mixing behaviour.

\begin{figure}[t!]
    \centering 
    \begin{subfigure}{0.32\textwidth}
        \centering \includegraphics[width=\columnwidth]{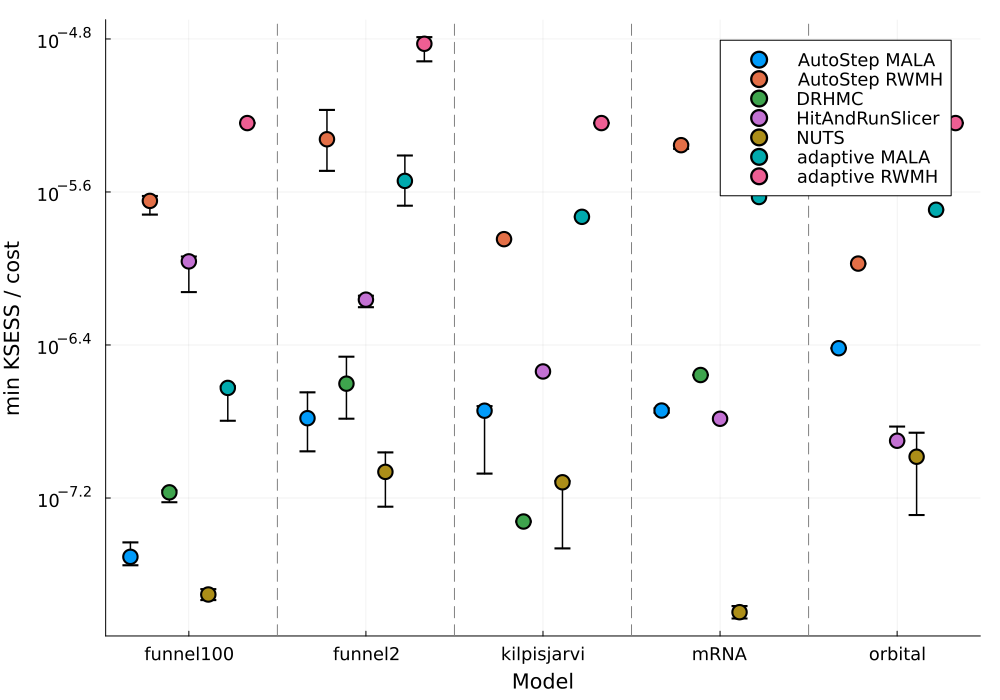}
        \caption{\texttt{minKSESS} per unitcost}
        \label{fig:minKSess_per_cost}
    \end{subfigure}
    \begin{subfigure}{0.32\textwidth}
        \centering \includegraphics[width=\columnwidth]{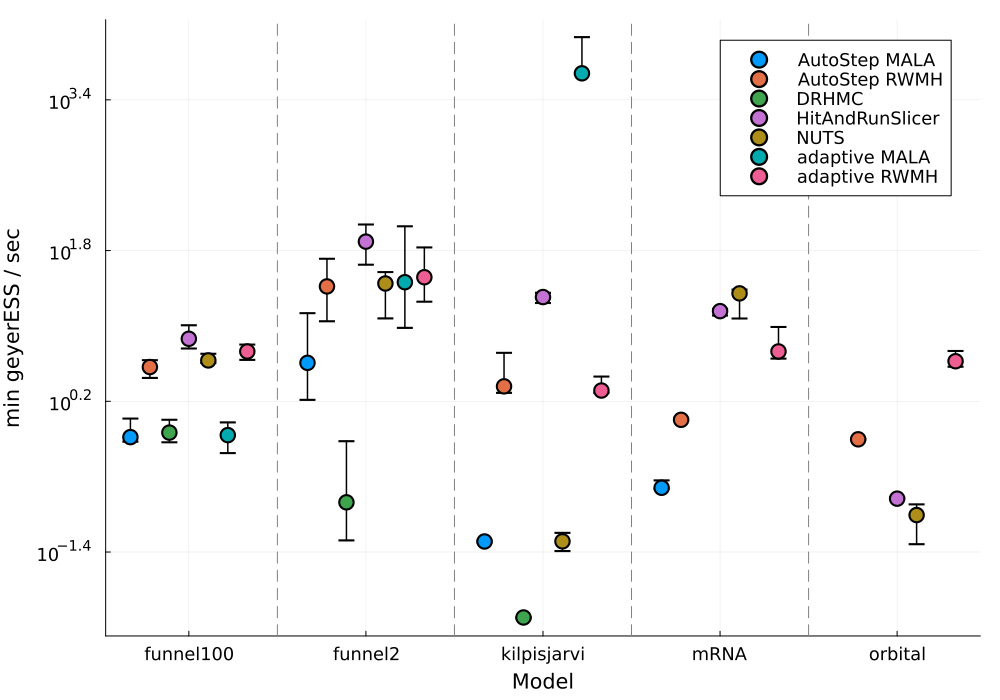}
        \caption{\texttt{minESS} per second}
        \label{fig:geyerESS_per_sec}
    \end{subfigure}
    \begin{subfigure}{0.32\textwidth}
        \centering \includegraphics[width=\columnwidth]{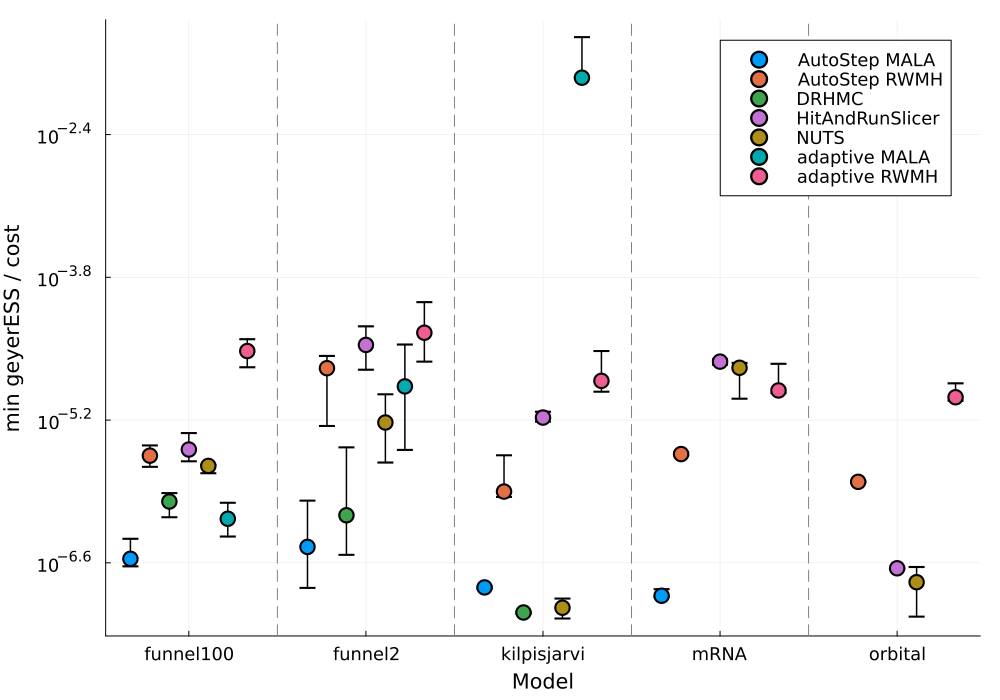}
        \caption{\texttt{minESS} per unit cost}
        \label{fig:geyerESS_per_cost}
    \end{subfigure}
    \caption{The same comparisons as \ref{fig:minKSess_per_sec}, except that we present \texttt{minKSESS} per unit cost (\cref{fig:minKSess_per_cost}), \texttt{minESS} per second (\cref{fig:geyerESS_per_sec}), and \texttt{minESS} per unit cost (\cref{fig:geyerESS_per_cost}), instead of \texttt{minKSESS} per second. }
    \label{fig:comp_existing}
\end{figure}

We also present additional results in \cref{fig:comp_existing} for the experiments 
in \cref{sec:experiment_compare_existing}. The \texttt{minESS} metrics were 
computed using the method described by \citet{Geyer92} in Section 3.3, which we 
found to be more reliable than the default settings in \texttt{MCMCChains}.

\section{Reference pair plots}

\begin{figure}[t]
\includegraphics[width=\textwidth]{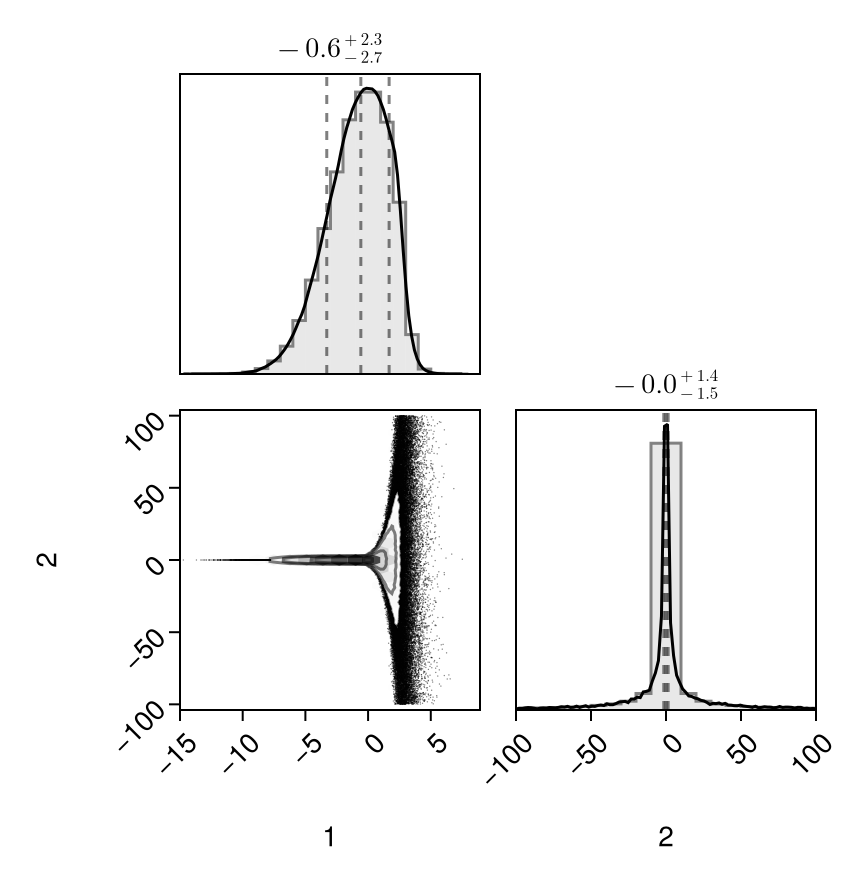}
\caption{Reference samples for \texttt{funnel2}.}\label{fig:reffunnel2}
\end{figure}

\begin{figure}[t]
\includegraphics[width=\textwidth]{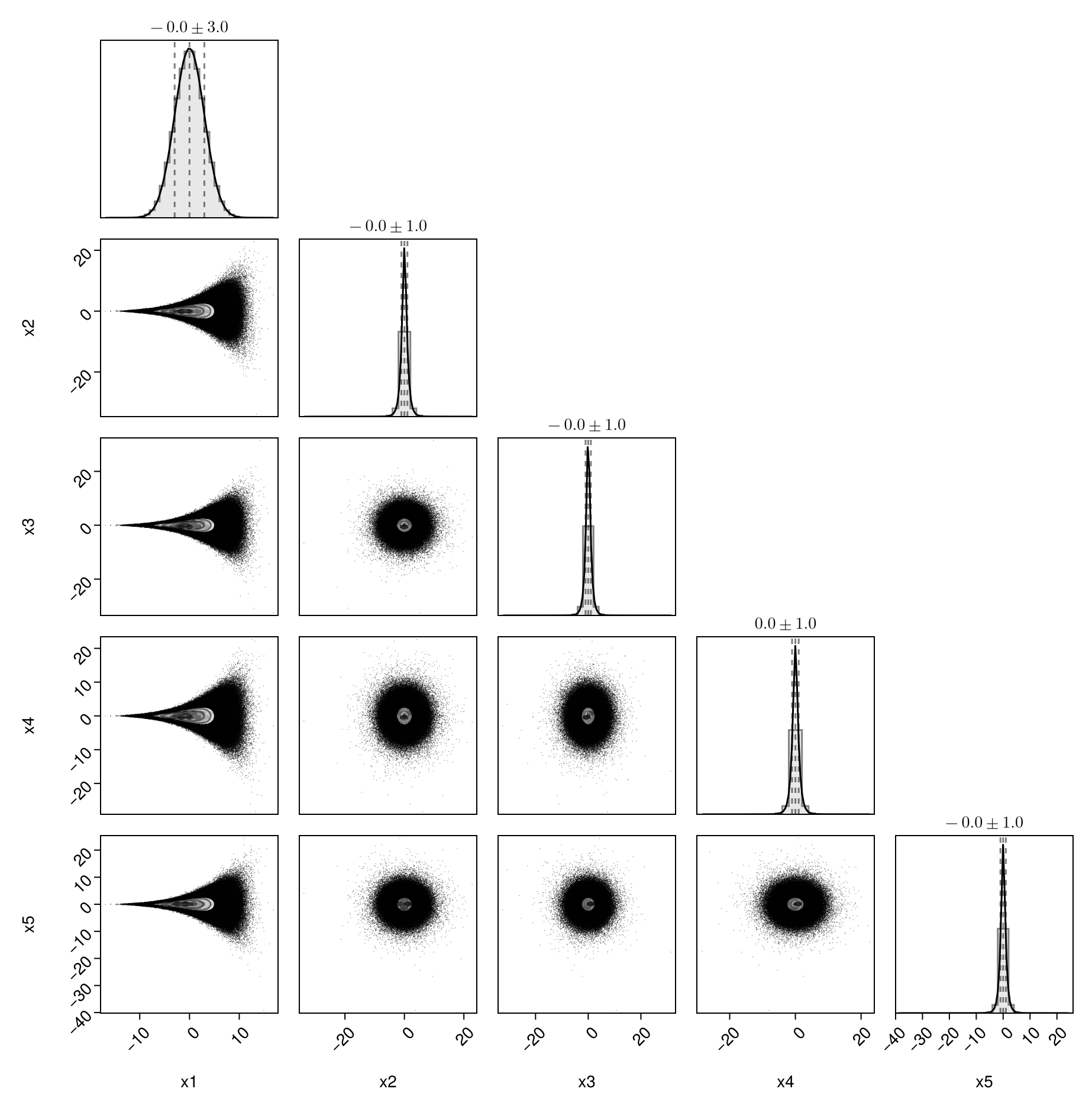}
\caption{First 5 dimensions of the reference samples for \texttt{funnel100}}\label{fig:reffunnel100}
\end{figure}

\begin{figure}[t]
\includegraphics[width=\textwidth]{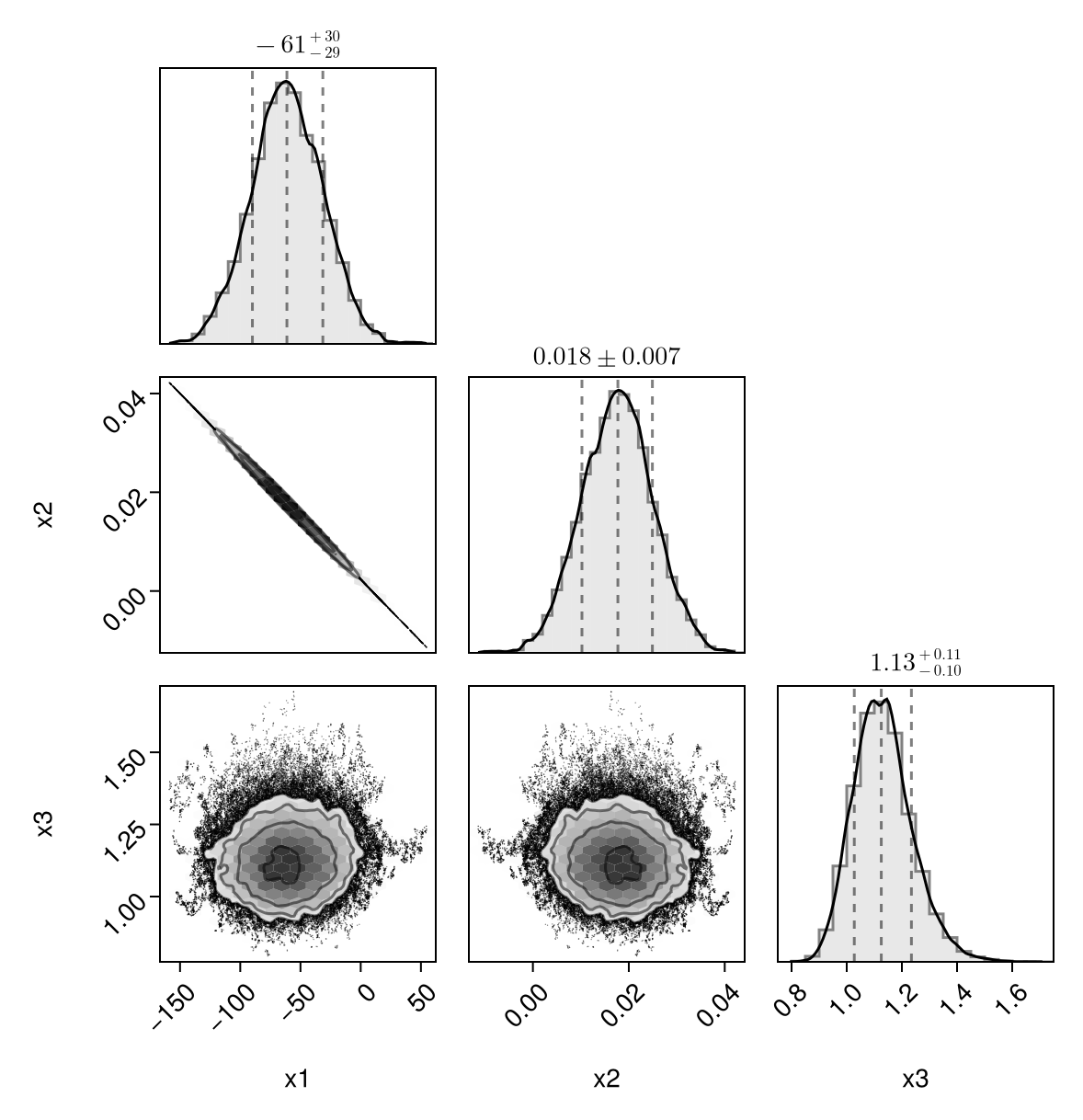}
\caption{Reference samples for \texttt{kilpisjarvi}.}\label{fig:refkilpis}
\end{figure}

\begin{figure}[t]
\includegraphics[width=\textwidth]{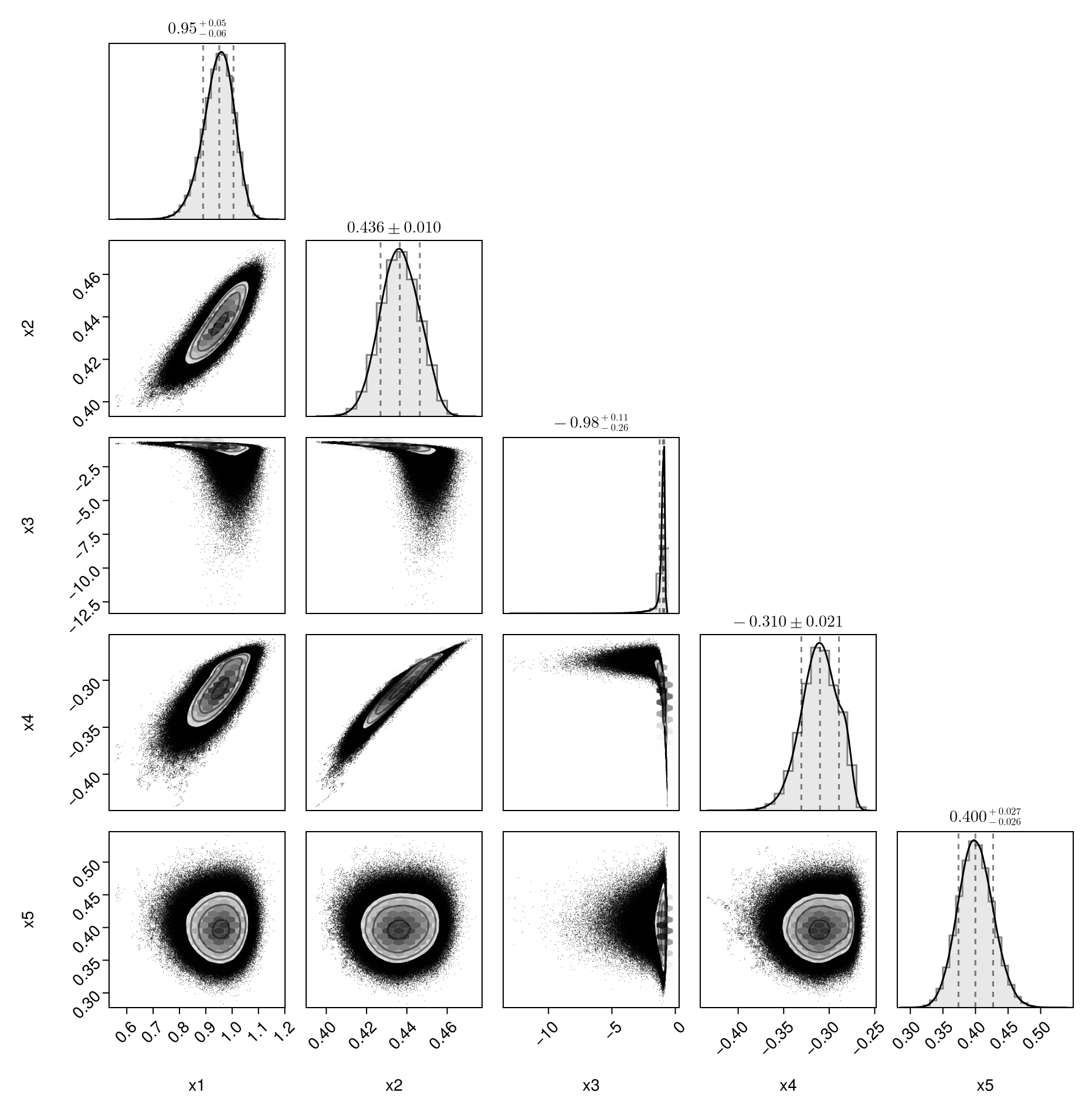}
\caption{Reference samples for \texttt{mRNA}.}\label{fig:refmrna}
\end{figure}

\begin{figure}[t]
\includegraphics[width=\textwidth]{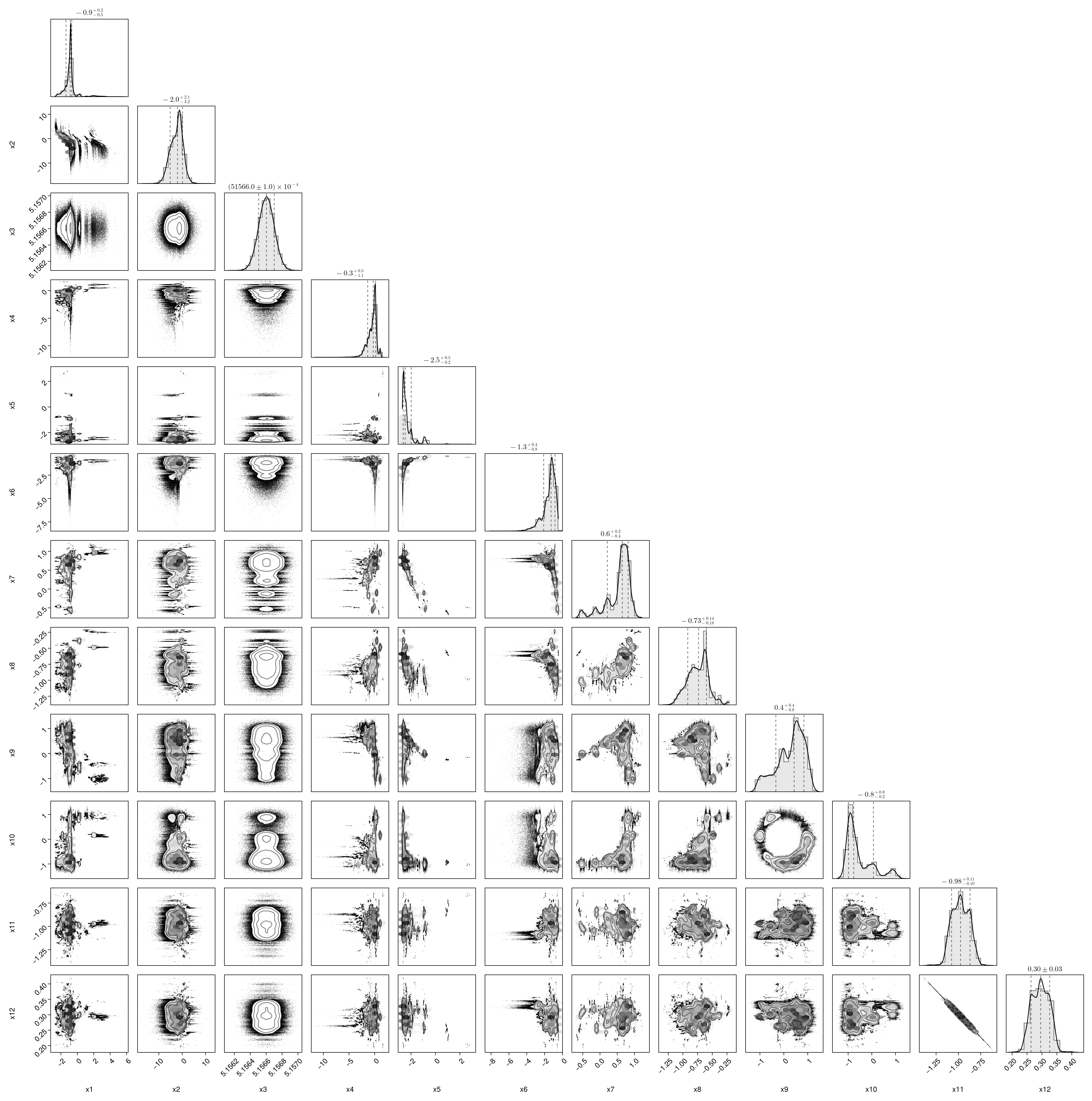}
\caption{Reference samples for \texttt{orbital}.}\label{fig:reforbital}
\end{figure}

\cref{fig:reffunnel2,fig:reffunnel100,fig:refkilpis,fig:refmrna,fig:reforbital}
display pair plots of the reference draws for each of the experiments in \cref{sec:experiment_compare_existing}.
Each set of reference draws contains between $10^6$--$10^7$ draws.
These draws were used for computing the \texttt{KSESS} estimates.

\section{\texttt{KSESS} diagnostic}\label{sec:ksess}

Suppose we obtain a collection of \iid draws $(x_t)_{t=1}^T$ from a target
distribution with known CDF $F(x)$ (or given a sufficiently large set of draws $(\stx_t)_{t=1}^{\stT}$, $\stT \gg T$ such that the empirical CDF of $(\stx_t)_{t=1}^{\stT}$ can be used in place of $F$).
For any $T\in\nats$, define
\[
	D((x_t)_{t=1}^T) = \sup_x \lt| F_T(x) - F(x) \rt|,
\]
where $F_T(x)$ is the empirical distribution of $(x_t)_{t=1}^T$.
It is known that $\sqrt{T} D((x_t)_{t=1}^T) \convd K$ as $T\to\infty$ where $K$ follows the Kolmogorov distribution \citep{Kolmogorov33} (see also \citet{Marsaglia03}),
\[
	\P\lt(K \leq x\rt) &= \frac{\sqrt{2\pi}}{x} \sum_{k=1}^\infty e^{-\frac{(2k-1)^2\pi^2}{8x^2}}\\
	\E[K] &= \log 2 \sqrt{\frac{\pi}{2}}\\
	\Var[K] &= \frac{\pi^2}{12} - \frac{\pi(\log2)^2}{2}.
\]
Heuristically, given a sample of size $T = N B$, we have that as $N, B\to\infty$,
\[
	\frac{1}{N}\sum_{\tau = 0}^{N-1}  \sqrt{B} D\lt((x_t)_{t=\tau B+1}^{(\tau+1)B}\rt) \approx \E[K],
\]
and hence the sample size $T$ is approximately
\[
	T &\approx \texttt{KSESS}_1 := T\lt(\frac{\log 2 \sqrt{\frac{\pi}{2}}}{\frac{1}{N}\sum_{\tau = 0}^{N-1} \sqrt{B} D\lt((x_t)_{t=\tau B+1}^{(\tau+1)B}\rt)}\rt)^2.
\]
If instead of exact Monte Carlo draws from the target, we obtain draws from a (potentially slowly mixing) Markov chain,
we expect the $D$ statistic to follow a convergence law roughly of the form $\sqrt{\alpha T} D((x_t)_{t=1}^T) \convd K$, where $\alpha T$
is the effective sample size. In this case,
\[
	\texttt{KSESS}_1 &= T\lt(\frac{\sqrt{\alpha}\log 2 \sqrt{\frac{\pi}{2}}}{\frac{1}{N}\sum_{\tau = 0}^{N-1} \sqrt{\alpha B} D\lt((x_t)_{t=\tau B+1}^{(\tau+1)B}\rt)}\rt)^2\\
	&\approx T\lt(\sqrt{\alpha}\rt)^2 = \alpha T,
\]
as desired. However, this doesn't measure severe failures well; note that the minimum possible 
value of $\texttt{KSESS}_1$ is $T \lt(\log 2 \sqrt{\frac{\pi}{2}}\rt)^2 / B$, which occurs when $D = 1$ for each batch (i.e., when the sampler
is working as poorly as possible). Hence we also consider the metric
\[
T\approx \texttt{KSESS}_2 &= \lt(\frac{\log 2 \sqrt{\frac{\pi}{2}}}{D((x_t)_{t=1}^T)}\rt)^2,
\]
which characterizes severe failure, but has too high variance to be useful when the sampler is functioning well.
Hence the final \texttt{KSESS} metric we use is
\[
\texttt{KSESS} = \lt\{\begin{array}{ll}
\texttt{KSESS}_2 & \texttt{KSESS}_2 \leq T\lt(\log 2 \sqrt{\frac{\pi}{2}}\rt)^2 / B\\
\texttt{KSESS}_1 & \text{otherwise}
\end{array}\rt.
\]
In our experiments we set $N = 40$.

\end{document}